\documentclass{article}
\usepackage{bm}
\usepackage{bbm}
\usepackage{xcolor}
\usepackage{pifont}
\usepackage{amsmath, amssymb}
\usepackage{amsthm}
\usepackage{mathtools}
\usepackage{nicematrix}
\usepackage{amssymb}
\usepackage{apacite}
\usepackage{natbib}
\usepackage{graphicx}
\usepackage{adjustbox}
\usepackage{caption}
\usepackage{lscape}
\usepackage{subcaption}
\usepackage{array}
\usepackage[linesnumbered,ruled,vlined]{algorithm2e}
\usepackage[utf8]{inputenc}
\usepackage[a4paper, total={6.5in, 9in}]{geometry}

\newsavebox{\mybox}\newsavebox{\mysim}
\newcommand{\distras}[1]{%
  \savebox{\mybox}{\hbox{\kern3pt$\scriptstyle#1$\kern3pt}}%
  \savebox{\mysim}{\hbox{$\sim$}}%
  \mathbin{\overset{#1}{\kern\z@\resizebox{\wd\mybox}{\ht\mysim}{$\sim$}}}%
}
\makeatother

\usepackage[colorlinks=true,citecolor=blue,allcolors=blue]{hyperref} 

\title{\textbf{Fast Meta-Analytic Approximations for Relational Event Models: Applications to Data Streams and Multilevel Data}}
\author{{Fabio Vieira}$^{1}$ \and Roger Leenders$^{2,3}$ \and Joris Mulder$^1$}
\date{%
    $^1$Department of Methodology and Statistics, Tilburg University\\%
    $^2$Department of Organization Studies, Tilburg University \\
    $^3$Jheronimus Academy of Data Science\\[2ex]%
    \today
}

\setcitestyle{aysep={}} 
\providecommand{\keywords}[1]
{
  \small	
  \textbf{Keywords:} #1
}

\newcolumntype{R}{@{\extracolsep{0.2cm}}r@{\extracolsep{10pt}}}%

\newcommand{\cmark}{\textcolor{green!80!black}{\ding{51}}}
\newcommand{\xmark}{\textcolor{red}{\ding{55}}}

\begin{document}

\RestyleAlgo{ruled}

\maketitle

\begin{abstract}
Large relational-event history data stemming from large networks are becoming increasingly available due to recent technological developments (e.g. digital communication, online databases, etc). This opens many new doors to learn about complex interaction behavior between actors in temporal social networks. The relational event model has become the gold standard for relational event history analysis. Currently however, the main bottleneck to fit relational events models is of computational nature in the form of memory storage limitations and computational complexity. Relational event models are therefore mainly used for relatively small data sets while larger, more interesting datasets, including multilevel data structures and relational event data streams, cannot be analyzed on standard desktop computers. This paper addresses this problem by developing approximation algorithms based on meta-analysis methods that can fit relational event models significantly faster while avoiding the computational issues. In particular, meta-analytic approximations are proposed for analyzing streams of relational event data and multilevel relational event data and potentially of combinations thereof. The accuracy and the statistical properties of the methods are assessed using numerical simulations. Furthermore, real-world data are used to illustrate the potential of the methodology to study social interaction behavior in an organizational network and interaction behavior among political actors. The algorithms are implemented in the publicly available R package 'remx'.
\end{abstract}


\section{Introduction}
\label{sec:intro}

\par
Social network analysis is the field that studies social structures by investigating human behavior with the goal of unveiling characteristics that affect actions and their consequences in social interactions \citep{scott1988social, wasserman1994social, brass2022new}. Recent technological developments (e.g., digital communication, online databases)  enable researchers to acquire richer and more extensive data about the social interactions between actors resulting in a more in-depth description of social interaction dynamics and improved predictions across various disciplines. Examples include the study of friendships \citep{goodreau2009}, social learning in Massive Open Online Courses \citep{vu2015relational}, the development of relations within teams \citep{leenders2016once}, inter-hospital patient transfers \citep{vu2017relational}, analysis of microstructures in financial networks \citep{ZAPPA2021125557}, social hierarchies \citep{redhead2022social}, the development of social relations among freshmen \citep{meijerink2023discovering}, and many others.

\par
In relational-event networks, the data consists of discrete instances of interactions among a finite set of actors in continuous time \citep{butts2017relational}. A collection of these interactions is referred to as a sequence of ``relational events". In this perspective, \cite{butts2008} was the first to propose the relational event model as a basic framework to model social actions. Since then, this model has been explored and expanded in multiple directions. For example, \cite{vu2011continuous} has proposed a model with time-varying parameters, \cite{perry2013point} utilized a partial likelihood approach to model the receiver given the sender, \cite{stadtfeld2017dynamic} and \cite{stadtfeld2017interactions} built upon this approach and introduced a model which can explain actors' preferences in social interactions, and \cite{mulder2019modeling} investigated the temporal evolution of networks by estimating a model in overlapping intervals.

\par
However, these models suffer from computational issues mostly associated with large data structures, which limit our ability to learn complex social interaction dynamics from larger data acquired using new technological developments. The data array that is used to fit these models usually has a $M \times D \times P$ dimensional structure, where $M$ is the number of events, $D$ is the number of dyads (i.e., directed pairs of actors), and $P$ is the number of predictor variables (e.g., endogenous statistics, exogenous statistics, or interactions thereof). Already in rather common situations would this data array run into storage problems. For example, if we want to learn about social interaction dynamics among colleagues using email traffic data from the Enron company (Klimt \& Yang, 2004), there are $M=32,261$ relational events (emails in this case), among $N=153$ actors, which correspond to $D=N(N-1)=23,256$ dyads. Furthermore, a typical relational event model would contain about $P=20$ predictors. The dimensions of the data array would then be $M \times D \times P=32,261\times 23,256\times 20$, which would contain approximately $1.5$e10 elements. On a regular desktop computer, however, the maximal memory capacity would already be reached after the first $M=4,000$ events. In another paper, \cite{brandes2009networks} considered four political networks where the number of countries/regions ($N$) varied from 200 to 700 countries/regions and relational event sequences of 20,000 to 171,000 political events. Thus, while such data dimensions are quite common in practice, they cannot be stored in memory using standard desktop computers. Therefore relational event models cannot be fitted to such data to learn about social interaction dynamics in temporal networks.

\par
In this paper, we provide a solution that allows relational event models to be fitted to such data formats on regular computers routinely. We shall focus on two different scenarios of relational-event data analyses. Firstly, we treat the case of streaming relational event data. A data stream consists of a set of observations that is continuously augmented by the arrival of new (batches of) data points \citep{ippel2019estimating, ippel2019online}. For instance, in \cite{brandes2009networks}, events are daily updated, whereas in \cite{boschee2015} the relational event sequence is weekly brought up to date. These examples explicit the need for modeling relational events in a real-time fashion. Secondly, we treat the case of relational event data with hierarchical structures using multilevel relational event models. \cite{betancourt2015hamiltonian} discuss in great length the issues associated with the fundamental structure of multilevel models and how parameter dependencies slow down model estimation. For relational event models, those problems add up to the fact that the alien form of the likelihood prevents modelers from taking full advantage of non-centered reparameterizations \citep{carpenter2017stan}. 

\par
The solution that we present extends methodological ideas from the meta-analysis literature to the relational event modeling framework. Meta-analyses are used to combine the information (estimates) from multiple scientific studies using specific summary statistics (e.g., estimates and error (co)variances) which were computed from the separate studies, to obtain accurate estimates on a global level \citep{raudenbush1985empirical, sutton2001bayesian, higgins2009re, borenstein2011introduction}. 
In the case of data streams where relational events are observed in an iterative manner, the batches of relational events are considered to be separate `studies' in the meta-analysis. Thus, relational event models are estimated separately per batch, and subsequently, the separate estimates and error covariance matrices are combined into one global estimate using a fixed-effect meta-analytic model. This fixed-effect model assumes that the effects are constant across studies \citep[e.g.][]{gronau2017bayesian} similar to a standard relational event model where it is assumed that the effect remains unchanged throughout the event stream. Thus, by treating subsequent batches of our relational event data stream as ``independent studies", the fixed-effect meta-analytic model can properly approximate the full relational event model based on the entire sequence. In the case of multilevel data, a random-effect meta-analytic model is used to approximate the multilevel relational-event model. The multilevel model is rooted in the assumption of the underlying existence of heterogeneity among the independent networks. Hence, the random-effects model is a natural choice to analyze these data \citep{van2009random}. Moreover, a mixed-effect model can be developed by combining the fixed- and random-effects meta-analytic models. Therefore, the multilevel model can be implemented by combining a structure identical to the fixed-effects model for data streams and a random-effects model. 

\par
Simulation study were conducted to assess how these approaches are capable of considerably reducing the computational time needed for fitting multilevel models in this class and, at the same time, providing results comparable to the exact model. Moreover, we develop a Gibbs sampler algorithm for the data stream case, where the avoidance of revisiting the past allows for faster inferential updates. The code developed to carry out the estimation of these models is available on GitHub (\url{https://github.com/TilburgNetworkGroup/remx}). The methodology that is presented in this paper on the other hand is implemented in the R package `remx'. An empirical analysis to study social interaction behavior using corporate email communications \citep{klimt2004introducing} and an empirical study of social interactions among political actors \citep{boschee2015} are performed to illustrate the models in action with real-world data.

\par
Finally it is important to note that other efforts have been made to improve the speed of fitting relational event models. The most notable one is using case-control sampling to avoid using the entire data matrix \citep{lerner2013modeling, vu2015relational}. In this methodology, besides the observed dyad for a given event, a subset of dyads is randomly drawn from the riskset that was not observed for each event. Their results indicated that such methods can result in accurate estimations. This method however has been particularly developed for the case of a single relational event sequence. It is yet unclear how to extend this sampling technique to the case of relational event data streams where new (batches of) relational events are pouring in on a regular basis or when considering multilevel relational event data. The proposed meta-analytic approximations on the other hand can straightforwardly be applied to the case of a single large relational event sequence. Furthermore, case-control sampling techniques are not readily available in existing R packages for temporal social network analysis. On the other hand, 'remx' is available as free R package.

\par
The remainder of this text is structured as follows: Section \ref{sec:rem} presents the relational event model framework; Section \ref{sec:comp_challenge} briefly discusses a few of its computational challenges; Section \ref{sec:remApp} introduces the meta-analytic approximations; Section \ref{sec:sim} contains the results of synthetic data studies; Section \ref{sec:applic} displays applications with empirical data; and Section \ref{sec:conclus} concludes the text by briefly restating the main points.


\section{Relational event modeling}
\label{sec:rem}
\subsection{Basic relational event models}
The relational event model is used to analyze a temporal social network represented by an ordered sequence of events, which are characterized by time-stamped interactions among a finite set of actors \citep{butts2008, butts2017relational}. We represent each event by $e = (s, r, t)$, where $t$ is the time when the interaction took place, $s$ and $r$ are indexes representing the actors who took part in this interaction. In a directed network, we call $s$ the sender and $r$ the receiver. Considering a network of $N$ actors, events are assumed to be randomly sampled from a set containing all pairs of actors that are at risk at $t$ (without loss of generality, in this paper we will assume all pairs of actors are at risk at each point in time). possible pairs of actors at every point in time. This set is called the risk set, $\mathcal{R}(t) = \{(s,r): s, r \in 1, 2, \dots, N\ |\ s \neq r\}$. Thus, assuming we observe $M$ events in a time window $[0, \tau)\ \in \mathbb{R}^{+}$, a relational event sequence is formally defined as
\begin{equation*}
    \textbf{E} = \{e_{m} = (s_m, r_m, t_m): (s_m, r_{m}) \in \mathcal{R}(t_{m});\ 0 < t_{1} < t_{2}, \dots, t_{M} < \tau  \},
\end{equation*}
\noindent
where $t_{m}$ is the time of the $m^{th}$ event, and $s_{m}$ and $r_{m}$ are the actors involved in the event at time $t_{m}$. The relational event model focuses on modeling the rates of interaction, $\lambda_{sr}(t)$, in the social network. \cite{butts2008} considers the inter-event times, $\Delta_{m} = t_{m} - t_{m-1}$, as exponentially distributed and assumes $\lambda_{sr}(t)$ to be constant between events, resulting in a piece-wise constant exponential model \citep{friedman1982piecewise}. The survival function of this model is given by $S_{sr}(\Delta_{m}) = \exp \{ - (t_{m} - t_{m-1}) \lambda_{sr}(t_m)\}$. Therefore, the likelihood has the following form
\begin{equation}
    \label{likelihood_butts}
    p_{time}(\textbf{E}\ |\ \bm{\beta}) = \prod_{m=1}^M  p_{time}(e_m\ |\ \bm{\beta}) = \prod_{m=1}^{M} \Big[ \lambda_{s_{m} r_{m}} (t_{m} | \textbf{E})\ \times \prod_{(s, r) \in \mathcal{R}(t_m)} S_{sr} (\Delta_{m} | \textbf{E}) \Big] \times \prod_{(s, r) \in \mathcal{R}(\tau)} S_{sr} (\Delta_{\tau} | \textbf{E})
\end{equation}
\noindent
where $\bm{\beta}$ is the parameter vector and $\bm{X}$ is a matrix with covariates. The rates of interaction are assumed to have a Cox regression form: $\lambda_{sr}(t|\textbf{E}) = \exp \{ \bm{x}'_{sr}(t)\bm{\beta}\}$ \citep{cox1972regression}, where $\bm{x}_{sr}(t)$ is a vector of $P$ predictor variables from the matrix $\bm{X}$ of the directional pair of sender $s$ and receiver $r$. The predictor variables can include exogenous variables (such as actors' attributes) and endogenous statistics, which summarize the event history until time $t$ which would be relevant to explain the social interaction between actors \citep[such as inertia, reciprocity, in-coming shared partners, turn-taking, etc.,][]{butts2008,leenders2016once}, as well as potential interactions between endogenous variables and exogenous variables. Note that these time-varying endogenous statistics need to be updated after each event. As social interaction behavior can be caused by complex social processes, relational event models may consist of many different potentially important predictors. For example, \cite{karimova2023separating} considered a relational event model with 103 predictor variables to explain and predict the voice loops during NASA's famous (but disastrous) Apollo 13 mission to the moon. Thus, it is important to note that the number of predictor variables (which may need to be updated after every event) can easily become very large.

In case only the order in which the events occurred is known (so the exact timing $t_m$ is not observed), the likelihood of the sequence of observed dyads $(s_m,r_m)$ becomes
\begin{equation}
    \label{likelihood_cox}
    p_{order}(\textbf{E}\ |\ \bm{\beta}) = \prod_{m=1}^M  p_{order}((s_m,r_m)\ |\ \bm{\beta}) = \prod_{m=1}^{M} \frac{\lambda_{s_m r_m}(t_{m} | \textbf{E})}{\sum_{(s,r) \in \mathcal{R}(t_m)} \lambda_{s r}(t_{m} | \textbf{E})}.
\end{equation}
Throughout this paper, when data come from the relational event model, we shall write this as follows
\[
\textbf{E} |\bm{\beta} \sim REM(\bm{\beta}).
\]
Depending on whether the exact timing is known, either the temporal or the ordinal REM can be used.

\noindent


Finally, it is important to note that both the ordinal and the temporal relational event models can be written as specific Poisson regression models (see also [REF]). This important property facilitates the classical estimation with software such as \textit{\textbf{glm}} in R. Details are provided in Appendix \ref{app:poisson}. Moreover, for Bayesian analysis of these models, researchers can make use of the R packages \textit{\textbf{relevent}} \citep{butts2008} and \textit{\textbf{remstimate}} \citep{Arena2022}, which are specifically designed for relational event analyses using endogenous network predictors \citep[see also][]{Meijerink2021}. Another option is the R package \textit{\textbf{rstanarm}} \citep{goodrich2022} but then users need to specify the model by hand. These Bayesian packages support a range of algorithms, such as Hamiltonian Monte Carlo to Bayesian importance sampling. Finally note that for a Bayesian analysis, a prior distribution needs to be specified for the coefficients. A common default choice is a noninformative flat prior \citep[see also][]{vieira2023bayesian}.

\subsection{Relational event models for data streams}\label{datastreams}

As relational-event data are often collected by means of digital communication applications, it is common that 
relational-event histories are sequentially collected in batches over time. 
We denote the $M_{\ell}$ events that were observed in batch $\ell$ according to $\textbf{E}(\ell)$, and the event sequence of all batches combined until batch $\ell$ as $\textbf{E}(1,\ldots,\ell)$. The likelihood until event batch $\ell$ can be written as a function of the likelihood of the data before batch $\ell$ combined with the likelihood of the new batch, i.e., 
\[
p(\textbf{E}(1,\ldots,\ell)|\bm\beta)
= p(\textbf{E}(\ell)|\bm\beta)\times p(\textbf{E}(1,\ldots,\ell-1)|\bm\beta),
\]
for $\ell=2,3,\ldots$, where the likelihood of event batch $\ell$ is equal to either \eqref{likelihood_butts} or \eqref{likelihood_cox} for the temporal REM or the ordinal REM, respectively, by substituting $\textbf{E}(\ell)$ for $\textbf{E}$. The goal is then to obtain the fitted model for all batches $\textbf{E}(1,\ldots,\ell)$ by updating the fitted model based on $\textbf{E}(1,\ldots,\ell-1)$ from the previous step with the newly observed batch $\textbf{E}(\ell)$. This would allow researchers to update the model on the fly as new batches of relational events are pouring in.

Even though it would be possible to set the batch size $M_{\ell}$ equal to the number of events that are recorded every time \citep[e.g., all events that are observed per day as in the data from][]{brandes2009networks}, it may be preferred to consider somewhat larger batches. There are two reasons for this. First, it would then need to update the model relatively often even though the estimates would remain largely unchanged. Second, when using the meta-analytic approximation, as will be discussed later, the events in one batch serve as one `study' from a meta-analytic perspective. From the meta-analytic literature is known that it is recommended to include studies of sufficient sample size such that the obtained estimates are reliable. For this reason, we will explore the influence of the batch size later in this paper in more detail.

As there are currently no R packages available specifically designed for relational event data streams, social network researchers need to rely on the same software as for a single relational event sequence, as was discussed in the previous section. Thus, every time a new batch of relational events is observed, the new events need to be combined with the entire sequence until that point, and the entire (updated) sequence can then be used for the estimation of the model.




\subsection{Multilevel relational event models}\label{multileveldata}

\par
Multilevel relational event models can be used to study the variation of network effects across different mutually independent clusters \citep{dubois2013hierarchical}. We will refer to each relational event history in this setting as an ``event cluster." The likelihood of the $K$ independent event clusters is defined as the probability of the events conditional on the cluster-specific effects, $\bm\beta_k$, and the common effects across clustered, denoted by $\bm\psi$, i.e., 
\begin{equation}
    \label{likelihood_butts_multi}
    p(\textbf{E}\ |\ \bm{\beta}) = \prod_{k=1}^{K} \Bigg[\prod_{m=1}^{M_{k}} \Big[ \lambda_{s_{m} r_{m}} (t_{m} | \textbf{E}_{k})\ \times \prod_{(s, r) \in \mathcal{R}_{k}(t_m)} S_{sr} (\Delta_{m} | \textbf{E}_{k}) \Big] \times \prod_{(s, r) \in \mathcal{R}_{k}(\tau_{k})} S_{sr} (\Delta_{\tau} | \textbf{E}_{k}) \Bigg]
\end{equation}
\noindent
where $\textbf{E} = \{\textbf{E}_{1}, \textbf{E}_{2}, \dots, \textbf{E}_{K}\}$, with $\textbf{E}_{k}$ the events in cluster $k$, and $M_{k}$, $\mathcal{R}_{k}(t)$ and $\tau_{k}$ are the number of events, the risk set and the end of the observation period for cluster $k$, for $k = 1, \dots, K$, respectively, and the rate parameter of events from actor $s$ towards actor $r$ can be written as $\lambda_{sr}(t|\textbf{E}_k) = \exp \{ \bm{z}'_{k,sr}(t)\bm{\psi} +  \bm{x}'_{k,sr}(t)\bm{\beta}_{k}\}$, where $\bm{z}_{k,sr}(t)$ is the vector of covariates of the common effects across clusters \citep[e.g., see][]{vieira2023bayesian}. In this context, it is assumed that the effects $\bm{\beta}_{k}$ in every cluster are independent and identically distributed across clusters,  following a normal distribution:
\begin{equation}
\label{hier_prior}
    \bm{\beta}_{k} \sim \mathcal{N}(\bm{\mu}, \bm{\Sigma}),\ \text{for}\ k = 1, 2, \dots, K,
\end{equation}
\noindent
where $\bm{\mu}$ is the mean effect in the population and $\bm{\Sigma}$ is the covariance matrix. Then, $\bm{\beta}_{k}$ contains the random effects, which are cluster-specific and sampled from its own population distribution of event cluster effects, similar as in a standard multilevel model \citep{gelman2006multilevel}. It is common to write the multilevel model as
\begin{eqnarray}
\label{multilevelREM}
\text{Level 1: }& \textbf{E}_k|\bm{\beta}_k,\bm\psi &\sim REM(\bm{\beta}_k,\bm\psi)\\
\nonumber \text{Level 2: }&\bm{\beta}_{k} &\sim \mathcal{N}(\bm{\mu}, \bm{\Sigma}),
\end{eqnarray}
where either the temporal or the ordinal model can be used on the first level depending on whether the exact timing is known. 
The main advantage of the multilevel structure is the automatic borrowing of information about network effects across event clusters. For instance, if an event cluster has a small sample size and thus there is little information to estimate the cluster-specific effect accurately, the estimate will be pooled towards the grand mean across all relational event clusters.

To fit multilevel relational event models, we can again make use of the Poisson regression formation of the model and thus use \textbf{lme4} \citep{bates2014fitting} for model fitting. Another possibility would be to use \textbf{rstanarm} for a Bayesian analysis. The computational burden of this method however is immense. For this reason \cite{vieira2022} considered only a subset of 15 classrooms of relational event sequences rather than the entire data set consisting of all 153 classrooms, which was not computationally feasible.

\section{Computational challenges}\label{sec:comp_challenge}

Currently, two crucial bottlenecks limit the applicability of REMs for routine use on personal computers to study temporal social networks: memory storage and computational complexity. Both aspects are discussed in this section.

\subsection{Memory storage}

\par
The risk set for event $m$ is the set that comprises all possible pairs of actors for which it would be possible to observe an event at every point time $m$. In the most general case, with a network of $N$ actors, the size of the risk set is equal to $N(N-1)$ in the case of directional relational events (and if all dyads are at risk). 
For instance, \cite{brandes2009networks} uses a data set with 202 political actors and 304,000 aggressive/cooperative events. They specified a model with 17 covariates. The risk set for this model has $202 \times(202 - 1) = 40,602$ possible dyads, so the data matrix is a 3-dimensional structure with dimensions 304,000 $\times$ 40,602 $\times$ 17. \cite{perry2013point} use a network of corporate email communications with 156 actors, 21,635 events, and 30 covariates. The size of the risk set is 24,180, resulting in a data matrix with dimensions 21635 $\times$ 24,180 $\times$ 30. \cite{lerner2013modeling} fit a relational event model to a data set with 168 actors belonging to ethnic groups or international organizations and 217,000 hostile/cooperative events. Their model contained 25 covariates. Thus, the data matrix has dimension 217,000 $\times$ 28,056 $\times$ 25. 

\par
The usual implementation of models with time-variant network statistics results in data objects that can easily become too large to be stored in the working memory of standard computers. This becomes even more problematic when a researcher has access to relational event data streams which potentially grow indefinitely. In data streams, the data size keeps growing over time intensifying the problem: as more data pours in, the data matrix grows larger and larger. For instance, the event data between countries and regions \citep{schrodt1994political}, which was considered by \cite{brandes2009networks}, are updated on a regular (e.g., daily) basis. Eventually, it becomes infeasible to store all predictor variables for all dyads for all events in memory in order to fit relational event models to the entire event sequence.

\par
In multilevel data, multiple networks are often analyzed in one step, which, again, requires storing many large data objects in memory at the same time. In this case, assuming one wants to estimate $P$ effects across $K$ networks consisting of $N_k$ actors each, we would need to store $\sum_{k=1}^K M_{k} \times N_{k}(N_{k}-1) \times P$ elements in working memory. In practice, where relational event data are collected using digital technologies, the memory would be drained on most personal computers and relational event model fitting would become unfeasible.

\subsection{Computational complexity}

\par
For single relational event histories, the number of computations that are necessary for the survival part of the likelihood in equation \eqref{likelihood_butts} and in the denominator in equation \eqref{likelihood_cox} grows approximately with the square of the number of actors. As a consequence, for moderately large networks, an enormous number of operations would be needed to compute the likelihood function. Therefore, if a researcher keeps observing social network interactions between the actors, this computational task easily becomes computationally infeasible. Moreover, given the software in R that is currently available, the relational event model would need to be estimated in one step based on the entire event history, and therefore the model would need to be re-estimated by updating the entire event sequence after every newly observed event (or batch), which would be quite problematic from a computational point of view. 

\par
For multilevel relational event data, computing the survival part of the likelihood function already requires $\sum_{k = 1}^{K} N_{k} (N_{k} - 1)$ operations (see equation \ref{likelihood_butts_multi}). Moreover, the cluster-specific effects are also tied together in a multidimensional structure represented by the population distribution. Hence, fitting this model would require the estimation of a large number of parameters. \cite{betancourt2015hamiltonian} discussed the computational challenges of this type of model, particularly with respect to the dependency among the parameters. Small changes in $\bm{\mu}$ and $\bm{\Sigma}$ result in drastic alterations in the population distribution, which makes model estimation a very complicated task.

\par
Hence, given the number of operations needed to compute the likelihood function and issues related to dependencies among parameters, fitting the relational event models previously described is a tedious and computationally expensive process. This is especially the case because most social network researchers do not have access to supercomputers but rely on their personal computers to perform their analyses, typically using R. Moreover, researchers who want to analyze large relational event data using a Bayesian perspective have even larger computational hurdles to overcome since the algorithms typically used for Bayesian analyses rely on iterative methods (e.g. Markov chain Monte Carlo methods) which usually require computing the likelihood a very large number of times.


\section{Meta-analytic approximations}
\label{sec:remApp}

In this Section, we present meta-analytic approximations to fit relational event models. These approximations can be used to fit models to relational event data streams, as well as a single large event sequence, and to multilevel relational event data. We will present these methods both from frequentist and Bayesian perspectives, showing that our approximation approach works for either modeling choice. Firstly, meta-analytic approximations are presented to estimate a relational event model for event streams in batches as described in Section \ref{datastreams}. Secondly, we discuss meta-analytic approximations for multilevel relational event models discussed in Section \ref{multileveldata}.

\subsection{Relational event data streams}
\label{meta_streaming}


\subsubsection{Frequentist meta-analytic approximation}\label{freqstream}



\noindent
Following the terminology of a meta-analysis, the observed relational events in the $\ell$-th batch can be viewed as a `study'. A relational event model can then be fitted to this $\ell$-th study resulting in a study specific estimate, $\hat{\bm{\beta}}({\ell})$, with error covariance matrix $\hat{\bm{\Omega}}({\ell})$.
Subsequently, the independent estimates are considered to be pseudo-data which can be pooled together in a fixed-effect meta-analytic model, i.e.,
\begin{equation}
\label{eq:fix_eff}
    \hat{\bm{\beta}}({\ell}) \sim \mathcal{N}(\bm{\beta}, \hat{\bm{\Omega}}({\ell})).
\end{equation}
In the meta-analysis literature, this setup is called a fixed effects meta-analytic model. Further, note that the normal approximation follows from large sample theory.
Now we consider the situation where $\ell$ batches were observed resulting in a pooled estimate which is denoted by $\tilde{\bm{\beta}}({\ell})$ and a multivariate Gaussian error covariance matrix $\tilde{\bm{\Omega}}({\ell})$. Next, the $(\ell+1)$-th batch is observed with approximate likelihood \eqref{eq:fix_eff}, and thus, following multivariate Gaussian theory, the updated estimate and error covariance matrix are given by
\begin{eqnarray}
\label{updateformula1}
\tilde{\bm{\Omega}}({\ell}+1) &=& (\hat{\bm{\Omega}}({\ell+1}))^{-1} +\tilde{\bm{\Omega}}({\ell}))^{-1})^{-1}\\
\tilde{\bm{\beta}}({\ell}+1) &=&
(\hat{\bm{\Omega}}({\ell+1}))^{-1} +\tilde{\bm{\Omega}}({\ell}))^{-1})^{-1}
(\hat{\bm{\Omega}}({\ell+1}))^{-1}\hat{\bm\beta}(\ell+1) +\tilde{\bm{\Omega}}({\ell})^{-1} \tilde{\bm\beta}(\ell))),
\end{eqnarray}
for $\ell=2,3,4,\ldots$, where $\tilde{\bm{\beta}}({1})=\hat{\bm{\beta}}({1})$ and $\tilde{\bm{\Omega}}(1)=\hat{\bm{\Omega}}(1)$. These formulas allow the updating of the estimates of the relational event coefficients and their uncertainty after every newly observed batch. It is easy to see that the mean vector and covariance matrix based on the first $\ell+1$ batches can also be written in the following non-iterative forms:
\begin{eqnarray}
\label{updateformula2}
\tilde{\bm{\Omega}}({\ell}+1) &=& \left(\sum_{l=1}^{\ell+1} \hat{\bm{\Omega}}(l)^{-1}\right)^{-1}\\
\tilde{\bm{\beta}}({\ell}+1) &=&
\left(\sum_{l=1}^{\ell+1} \hat{\bm{\Omega}}(l)^{-1}\right)^{-1}\left(\sum_{l=1}^{\ell+1} \hat{\bm{\Omega}}(l)^{-1}\hat{\bm\beta}(l)\right)
\end{eqnarray}

%
%
%
The steps to update the relational event model in the case of observing batches of events in a streaming setup are summarized in Algorithm \ref{alg:meta_freq_single}. We developed code for Algorithm \ref{alg:meta_freq_single}, which is available on the \textit{\textbf{remx}} package. 

\subsubsection{Bayesian meta-analytic approximation}
Here we discuss a Bayesian meta-analytic implementation of the relational event model when observing streams of event batches over time. The Bayesian method requires the assignment of a prior distribution, denoted by $p(\bm{\beta})$. The posterior of the vector of parameters based on the $\ell$-th batch can be obtained using Bayes theorem, i.e.,
\begin{equation}
    p (\bm{\beta}\ |\ \textbf{E}(\ell) ) = \frac{p ( \textbf{E}(\ell) | \bm{\beta} ) p (\bm{\beta})}{\int p ( \textbf{E}(\ell) | \bm{\beta} ) p (\bm{\beta}) d \bm{\beta}}
\end{equation}
\noindent
Samples from the posterior distribution can be obtained via simulation techniques (e.g., through Markov chain Monte Carlo). The Bayesian approach to relational event analysis suffers from the same issues as the frequentist approach but it has the added disadvantage of being considerably slower (refer to that section where you discuss this).

We propose the following approximation Bayesian solution. We shall use multivariate normal prior for the coefficients, denoted by $\bm\beta\sim \mathcal{N}(\bm\mu_0,\bm\Sigma_0)$. As a default setup, a noninformative flat prior can be used which can be constructed using a diagonal covariance matrix with huge diagonal elements together with zero means. In this case, the posterior would be completely determined by the likelihood. Furthermore, following Bayesian large sample theory, Gaussian approximations are again used to approximate the posterior based on every batch. The posterior based on the first batch can then be written as
\begin{eqnarray*}
p(\bm\beta|\textbf{E}(1)) \propto p(\textbf{E}(1)|\bm\beta)\times p(\bm\beta)
& \approx & \mathcal{N}(\bm{\beta};\bar{\bm{\beta}}({1}), \bar{\bm{\Omega}}({1})),\\
\text{with }~~
\bar{\bm{\Omega}}({1}) &=& \left(\hat{\bm{\Omega}}({1})^{-1}+\bm\Sigma_0^{-1}\right)^{-1}\\
 \bar{\bm{\beta}}({1}) &=&
\left(\hat{\bm{\Omega}}({1})^{-1}+\bm\Sigma_0^{-1}\right)^{-1}
\left(\hat{\bm{\Omega}}({1})^{-1}\hat{\bm\beta}(1)+\bm\Sigma_0^{-1}\bm\beta_0\right)
\end{eqnarray*}
where $\mathcal{N}(\bm{\beta};\bar{\bm{\beta}}, \bar{\bm{\Omega}})$ denotes a Gaussian distribution for $\bm{\beta}$ with mean vector $\bar{\bm{\beta}}$ and covariance matrix $\bar{\bm{\Omega}}$.
The concept of Bayesian updating, where the current posterior serves as the prior when observing new data, can directly be applied to the setting of relational event data streams. Mathematically, the update can be written as follows:
\begin{eqnarray*}
    p (\bm{\beta}\ |\ \textbf{E}(1,\ldots,\ell+1) )& \propto &p ( \textbf{E}(\ell+1) | \bm{\beta} ) p (\bm{\beta}\ |\ \textbf{E}(1,\ldots,\ell) )\\
    &\approx & \mathcal{N}(\bm{\beta};\hat{\bm{\beta}}({\ell}), \hat{\bm{\Omega}}({\ell}))
    \times
    \mathcal{N}(\bm{\beta};\bar{\bm{\beta}}({\ell}), \bar{\bm{\Omega}}({\ell}))\\
    &\approx &
    \mathcal{N}(\bm{\beta};\bar{\bm{\beta}}({\ell+1}), \bar{\bm{\Omega}}({\ell+1})),
\end{eqnarray*}
for $\ell=1,2,3,\ldots$, where the formulas for the mean and covariance matrix can be found in Algorithm \ref{alg:meta_freq_single}. Again the noniterative formulas for the posterior mean and covariance matrix can then be written as
\begin{eqnarray*}
\bar{\bm{\Omega}}({\ell+1}) &=& \left(\bm\Sigma_0^{-1}+\sum_{l=1}^{\ell+1}\hat{\bm{\Omega}}({l})^{-1}\right)^{-1}\\
 \bar{\bm{\beta}}({\ell+1}) &=&
\left(\bm\Sigma_0^{-1}+\sum_{l=1}^{\ell+1}\hat{\bm{\Omega}}({l})^{-1}+\right)^{-1}
\left(\bm\Sigma_0^{-1}\bm\beta_0+\sum_{l=1}^{\ell+1}\hat{\bm{\Omega}}({l})^{-1}\hat{\bm\beta}(l)\right)
\end{eqnarray*}
It is interesting to note that the resulting meta-analytic approximations based on a Bayesian approach are equivalent to the classical counterpart (Section \ref{freqstream}) when a noninformative prior would be used in the Bayesian approach.


%

\subsection{Multilevel relational event data}
\label{meta_multi}

\par
Let us now turn to the case where $K$ independent networks have been observed. The following methods can be applied to perform approximate multilevel analyses of relational event history data. Unlike the fixed effects meta-analytic method for the streaming data scenario, for multilevel relational event data, we propose a random effects meta-analytic method which allows the variability of the coefficients across event clusters and the borrowing of information across clusters, in the same spirit as in ordinary multilevel models. 



\subsubsection{Classical multilevel meta-analytic approximation}

\par
Following the terminology of a meta-analysis, the $k$-th event cluster now serves as the $k$-th `study'. Similar to meta-analyses, a relational event model is first fitted to each event cluster, and large sample theory is used to obtain a multivariate normal approximation of the effects in event cluster $k$. Subsequently, the meta-analytic mixed-effects approximation of the multilevel relational event model in \eqref{multilevelREM} can be written as
\begin{equation} 
\begin{gathered}
\label{eq:mixed_meta}
    \hat{\bm{\beta}}_{k}  \sim \mathcal{N} \Big(
    \bm{\mu}_{\beta} + \bm{\delta}_{k}, \hat{\bm{\Omega}}_{k} \Big)\\
    \bm{\delta}_{k} \sim \mathcal{N}(\bm{0}, \bm{\Sigma}), \text{ for $k=1,\ldots,K$},
\end{gathered}
\end{equation}
\noindent
where $\hat{\bm{\beta}}_{k}$ is the maximum-likelihood estimate of the coefficients in cluster $k$ and $\hat{\bm\Omega}_k$ is the error covariance matrix of the coefficients in cluster $k$.
Thus, the parameters to be estimated are $\bm{\delta}_{k}$, $\bm{\mu}_{\beta}$, $\bm{\Sigma}$, where $\bm{\mu}_{\beta}$ is the vector of fixed effects (which are common across all clusters),  $\bm{\delta}_{k}$ are the random-effects for group $k$ (which quantify the cluster-specific deviations from the fixed effects), and $\bm{\Sigma}$ is the random-effect covariance matrix which quantifies the (co)variability of cluster-specific deviations across clusters. 
The estimators for the random-effect parameters are given by

\begin{equation}
\begin{gathered}
    \bar{\bm{\delta}}_{k} = {\Big(\hat{\bm{\Omega}}^{-1}_{k} +  \bm{\Sigma}^{-1} \Big)}^{-1} \Big(\hat{\bm{\Omega}}^{-1}_{k} ( \hat{\bm{\beta}}_{k} - \bm{\mu}_{\beta}) \Big)\\
    \bar{\bm{\mu}}_{\beta} = \Big(\sum_{k=1}^{K} \hat{\bm{\Omega}}^{-1}_{k} \Big)^{-1} \Big( \sum_{k=1}^{K} \hat{\bm{\Omega}}^{-1}_{k} ( \hat{\bm{\beta}}_{k} - \bm{\delta}_{k}) \Big)\\
    \bar{\bm{\Sigma}} = \frac{1}{K} \sum_{k=1}^{K} \bm{\delta}_{k} \bm{\delta}'_{k}.
\end{gathered}
\end{equation}

\noindent
where $\bar{\bm{\delta}}_{k}$, $\bar{\bm{\mu}}_{\beta}$, and $\bar{\bm{\Sigma}}$ are the multilevel estimators. Then, an optimization method, such as the Newton-Raphson algorithm, can be used to obtain those estimates \citep{ypma1995historical}. An important property of $\bar{\bm{\delta}}_{k}$ concerns the shrinkage effect, which pulls the independent estimate towards the grand mean $\bm{\mu}_{\beta}$ in the case of considerable uncertainty about the group-specific estimate in the error covariance matrix. Therefore, clusters can borrow strength from other event clusters to obtain a shrunken estimate in the multilevel step. This is similar to the James-Stein estimator, which has been proven to be more efficient than the independent maximum-likelihood estimator when the data are nested in event clusters \citep{efron1977stein, james1992estimation}. This model can be fitted with the R packages \textit{\textbf{metafor}} \citep{viechtbauer2010} and \textit{\textbf{mixmeta}} \citep{gasparrini2019}.



\noindent

\subsubsection{Bayesian multilevel meta-analytic approximation}

The model described in the previous subsection is sufficient when it is realistic to assume that all coefficients vary across clusters. However, there are cases in multilevel relational event analyses where some coefficients are not heterogeneous across clusters but in fact, are constant across clusters \citep{vieira2022}. To accommodate that, the complete heterogeneous random effects meta-analytic model in equation \eqref{eq:mixed_meta} can be generalized to 

\begin{equation}
\begin{gathered}
    \begin{bmatrix}
        \hat{\bm{\psi}}_{k} \\  
        \hat{\bm{\beta}}_{k} 
    \end{bmatrix} \sim \mathcal{N} \Bigg(
    \begin{bmatrix}
        \bm{\psi} \\
        \bm{\mu}_{\beta}
    \end{bmatrix} + 
    \begin{bmatrix}
        \bm{0} \\
        \bm{\delta}_{k}
    \end{bmatrix},
    \begin{bmatrix}
        \hat{\bm{\Omega}}_{k,11} \ \ \ \ \ \ \ \ \ \ \\
        \hat{\bm{\Omega}}_{k,21} \ \ \ \hat{\bm{\Omega}}_{k,22} 
    \end{bmatrix}
    \Bigg)\\
    \bm{\delta}_{k} \sim \mathcal{N}(\bm{0}, \bm{\Sigma}).
\end{gathered}
\end{equation}

\noindent

A Bayesian MCMC algorithm is proposed for fitting this mixed-effects meta-analytic model. A uniform prior is specified for the joint distribution of $\bm{\psi}$ and $\bm{\mu}_{\beta}$, i.e., $p(\bm{\psi}, \bm{\mu}_{\beta}) \propto 1$. Next, to sample $\bm{\delta}_{k}$, we have that

\begin{equation}
    \begin{bmatrix}
        \bm{0} \\
        \bm{\delta}_{k}
    \end{bmatrix} \sim
    \mathcal{N} \Bigg(
    \begin{bmatrix}
        \hat{\bm{\psi}}_k - \bm{\psi}\\
        \hat{\bm{\beta}}_{k} - \bm{\mu}_{\beta}
    \end{bmatrix},
     \begin{bmatrix}
        \hat{\bm{\Omega}}_{k,11} \ \ \ \ \ \ \ \ \ \ \\
        \hat{\bm{\Omega}}_{k,21} \ \ \ \hat{\bm{\Omega}}_{k,22} 
    \end{bmatrix}
    \Bigg).
\end{equation}

\noindent
Hence, the conditional distribution of $\bm{\delta}_{k}$ is given by

\begin{equation}
    \begin{gathered}
        \bm{\delta}_{k} \sim \mathcal{N} \Big((\hat{\bm{\beta}}_{k} - \bm{\mu}_{\beta}) + \hat{\bm{\Omega}}_{k,12} {\hat{\bm{\Omega}}}^{-1}_{k,11} (\bm{0} - (\hat{\bm{\psi}}_{k} - \bm{\psi})),  \hat{\bm{\Omega}}_{k,22} - \hat{\bm{\Omega}}_{k,21} {\hat{\bm{\Omega}}}^{-1}_{k,11} \hat{\bm{\Omega}}_{k,12}  \Big).
    \end{gathered}
\end{equation}

\par
Finally, following \cite{huang2013simple}, we define a matrix Half-$t$ prior for the covariance matrix $\bm{\Sigma}$. This prior is specifically designed for random effects covariance matrices and is preferred over the Inverse-Wishart, because the Inverse-Wishart induces a degree of informativeness that affects posterior inferences \citep{gelman2006prior}. The matrix Half-$t$ on the other hand, induces Half-$t$ priors on the standard deviations in the diagonal of the covariance matrix. In addition, for a specific hyperparameter choice, it results in uniform priors between $(-1, 1)$ for the correlations between the random effects. Thus, this distribution enables non-informative priors for all standard deviations and correlation parameters. This prior is given by

\begin{equation}
\begin{gathered}
    \bm{\Sigma}\ |\ \textbf{A} \sim \text{Inv-Wishart} \big(\eta + P - 1, 2 \eta \textbf{A} \big) \\
    \textbf{A} = \text{diag} \big(1/\alpha_{1}, 1/\alpha_{2}, \dots, 1/\alpha_{P} \big) \\
    \alpha_{i} \sim \text{Inv-Gamma} \Big(\frac{1}{2}, \frac{1}{d^2_{i}} \Big),\ \text{for}\ i = 1, 2, \dots, P.
\end{gathered}
\end{equation}

\noindent
In the above expression, $\text{diag} \big(1/\alpha_{1}, 1/\alpha_{2}, \dots, 1/\alpha_{P} \big)$ represents a diagonal matrix with the values $(1/\alpha_{1}$, $1/\alpha_{2}$, $\dots$, $1/\alpha_{P}$). Furthermore, positive real values are set for $\eta \in \mathbb{R}^+$ and $d_{i} \in \mathbb{R}^+$, for $i = 1, 2, \dots, P$. Setting $\eta = 2$ leads to $\mathcal{U}(-1,1)$ over all correlation parameters. Moreover, as in \cite{gelman2006prior}, large values of $d_{i}$ result in vague prior distributions on the standard deviations. An advantage of this approach is that the Bayesian multilevel model is less prone to the degeneracy issues that have plagued the classic multilevel model, such as $\bm{\Sigma}$ being non-positive definite on the boundary of the parameter space \citep{chung2015weakly}. Algorithm \ref{alg:multirem} details the steps to estimate this model. The MCMC algorithm is implemented in the new R package \textit{\textbf{remx}}.


\section{Synthetic data studies}
\label{sec:sim}

\par
In this section, we examine the performance of the approximation algorithms.  Our objective is to investigate (i) the accuracy of the estimates as a function of the batch or cluster size (in comparison to the exact models), (ii) the computation time as a function of the sample size, and (iii) the efficiency of the multilevel estimators.

\subsection{Data streams}

\par
We generated a relational event history with $N = 25$ actors, $M = 5000$ events, and $P = 14$ effects: $10$ statistics (inertia ($\beta_{\text{inertia}} = 0.01$), reciprocity ($\beta_{\text{reciprocity}} = 0.01$), in-degree receiver ($\beta_{\text{indRec}} = -0.02$), in-degree sender ($\beta_{\text{indSnd}} = -0.02$), out-degree receiver ($\beta_{\text{outRec}} = -0.02$), out-degree sender ($\beta_{\text{indSnd}} = -0.01$), outgoing two path ($\beta_{\text{otp}} = 0.02$), incoming two path ($\beta_{\text{itp}} = -0.02$), outgoing shared partners ($\beta_{\text{osp}} = 0.01$), incoming shared partners ($\beta_{\text{isp}} = -0.01$)), $4$ interactions (inertia $\times$ reciprocity ($\beta_{\text{intXRecp}} = -0.01$), in-degree receiver $\times$ in-degree sender ($\beta_{\text{indRecXindSnd}} = -0.05$), out-degree receiver $\times$ out-degree sender ($\beta_{\text{outRecXoutSnd}} = -0.02$), outgoing two path $\times$ incoming two path ($\beta_{\text{otpXitp}} = -0.01$)) plus an intercept ($\beta_{\text{intercept}} = -5$). In this experiment, we emulated the desired streaming effect by gradually increasing the number of events in the network. However, the same framework can be applied to large networks. The increments were made in batches of 30, 50, 100, 150, 200, 300, and 500 events until the maximum of 5000 was reached. The objective was to compare the results from the meta-analytic approximations discussed in Subsection \ref{meta_streaming} with the exact model. Therefore, every new batch would augment the sequence of relational events constituting a new partition for the meta-analytic approximation and an entirely new sequence for the exact model. For model fitting, we used the R packages \textit{\textbf{remstimate}} for the exact model \citep{Arena2022}, \textit{\textbf{metafor}} for the frequentist meta-analytic approximation \citep{viechtbauer2010}, and \textit{\textbf{rstan}} for both the exact and approximate Bayesian method \citep{carpenter2017stan}.

\begin{figure}
    \centering
    \subfloat{\includegraphics[width=5in]{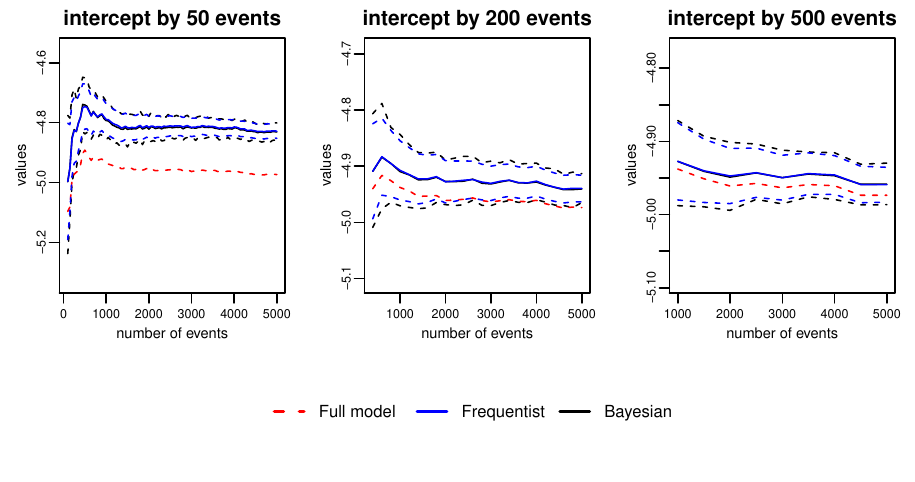}}\\ 
    \subfloat{\includegraphics[width = 5in]{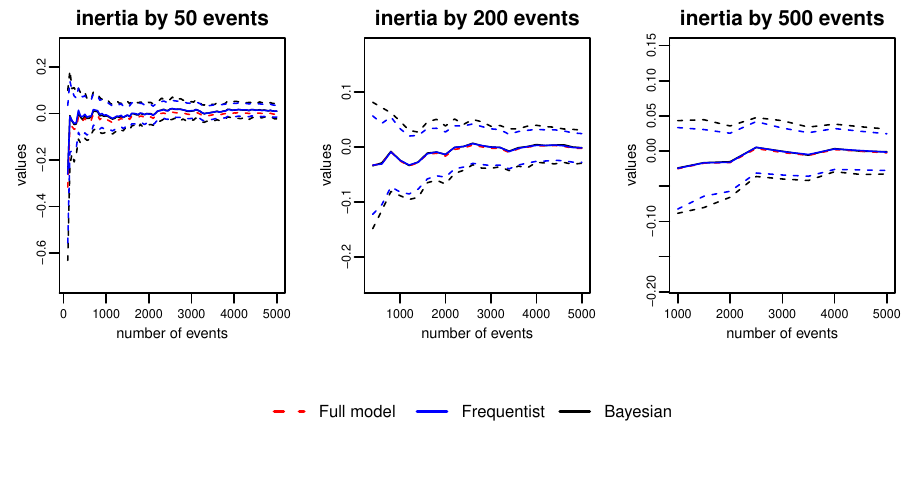}}\\
    \subfloat{\includegraphics[width = 5in]{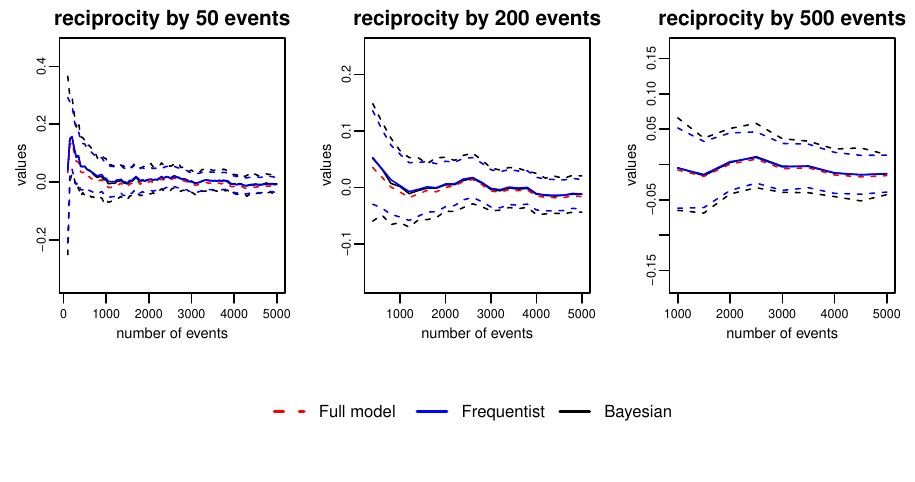}}
    \caption{Comparison of parameter estimates of the fixed-effect model and the exact model. The original sequences started with 100, 400 and 1000 events, respectively, and were incremented in batches of 50, 200 and 500. Dotted blue and black lines are 95\% intervals.}
    \label{fig:MLEFixed}
\end{figure}

\paragraph{Parameter recovery:} The estimation process starts with two batches, for all batch sizes. Figure \ref{fig:MLEFixed} shows the comparison between models for increments of 50, 200, and 500 events. The red, blue, and black lines represent, respectively, the exact model, the frequentist approximation, and the Bayesian approximation. For inertia and reciprocity, the size of the batch makes little difference and the estimates of the approximations are always close to the exact model. The same behavior is observed for all the statistics used in this simulation and is therefore omitted to keep the presentation of the results as concise as possible. The intercept, however, seems to absorb the bias when the size of the batch is small, (e.g. 50 events). The estimates of the meta-analytic approximations move in the direction of the exact model as the size of partitions increases.

\begin{figure}
    \centering{
    \subfloat{\includegraphics[width = 6.5in]{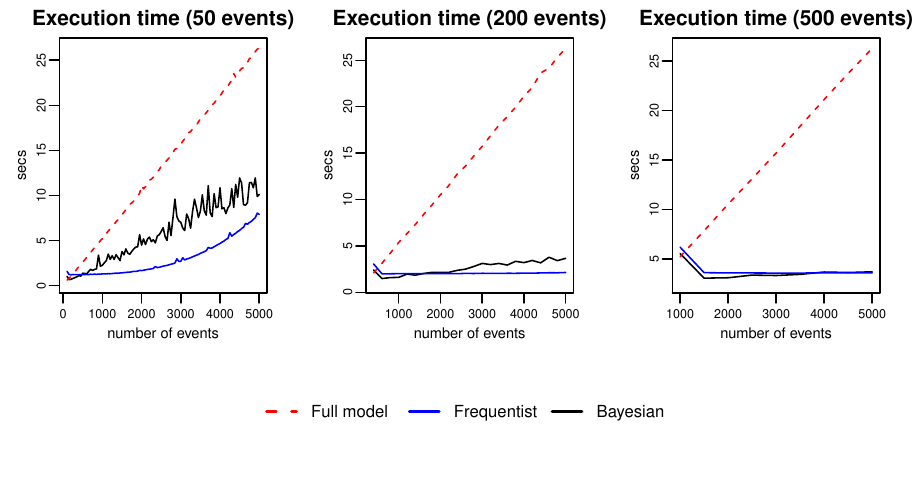}}
    }
    \caption{Comparison of running times for fixed-effect model. Sequences were incremented with batches of 50, 200 and 500 events.}
    \label{fig:run_time_fixEff}
\end{figure}

\paragraph{Computational time:} Finally, Figure \ref{fig:run_time_fixEff} shows the running times for each model. The exact model (dotted red line) is the only one that displays an upward linear trend as the number of events grows, due to the fact that the whole model needs to be re-estimated as new events are observed. Whereas, the meta-analytic approximation requires only an estimate for the new batch and then the pooling in the fixed-effect model. Thus, this reinforces the idea that eventually it will be infeasible to fit the exact model as the sequence grows larger and larger.

\subsection{Multilevel models}

\par
In this experiment, $K = 30$ independent networks were generated. The number of events was gradually increased in each network from $M_{k} = 50$ up to $M_{k} = 5000$, for $k = 1, \dots, K$. Then, $P = 6$ effects (inertia, reciprocity, outgoing two paths, incoming two paths, psABXA, psABXB) plus an intercept were fitted to those event clusters using the multilevel models described in Subsection \ref{meta_multi}. The goal was to check whether the approximations performed in a similar manner to the exact models. The Bayesian exact model and approximation were fit using \textit{\textbf{rstan}} \cite{carpenter2017stan}. The exact frequentist model was fit using \textit{\textbf{lme4}} \citep{bates2014fitting}, writing it as a Poisson regression (see Section \ref{sec:rem}). The frequentist approximation was fit with \textit{\textbf{metafor}} \citep{viechtbauer2010}. We evaluate the performance of these models along four different criteria: recovery of mean effects, the strength of shrinkage, the efficiency of the estimator, and their computational time.

\begin{figure}
    \centering{
    \subfloat{\includegraphics[width = 6.5in]{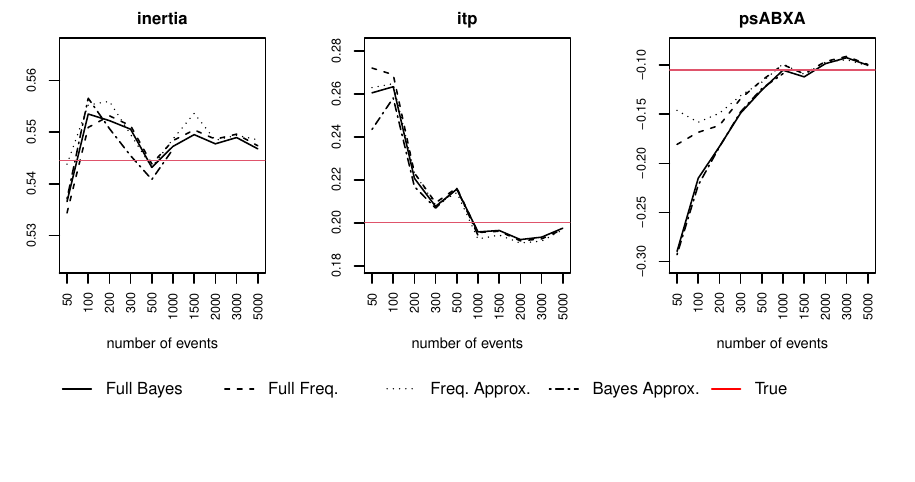}}\\ 
    }
    \caption{Comparison of recovery of mean effects for three different covariates.}
    \label{fig:Mean_eff}
\end{figure}

\paragraph{Recovery of mean effects:} We compare the estimators of the four models to see how well they recover the true mean effect that generated our data sets. Figure \ref{fig:Mean_eff} shows the recovery for three covariates across different sample sizes. There is a difference between the estimators for the approximation and the exact models for small sample sizes. However, they clearly converge as the number of events grows, going in the direction of the true mean effect.

\begin{figure}
    \centering{
    \subfloat{\includegraphics[width = 6.5in]{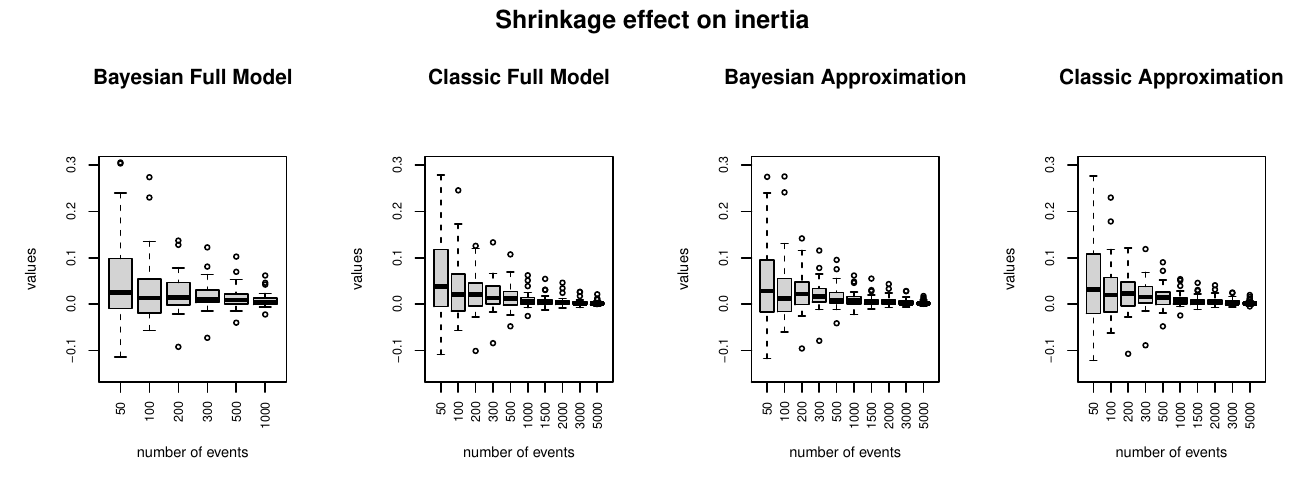}}\\ 
    \subfloat{\includegraphics[width = 6.5in]{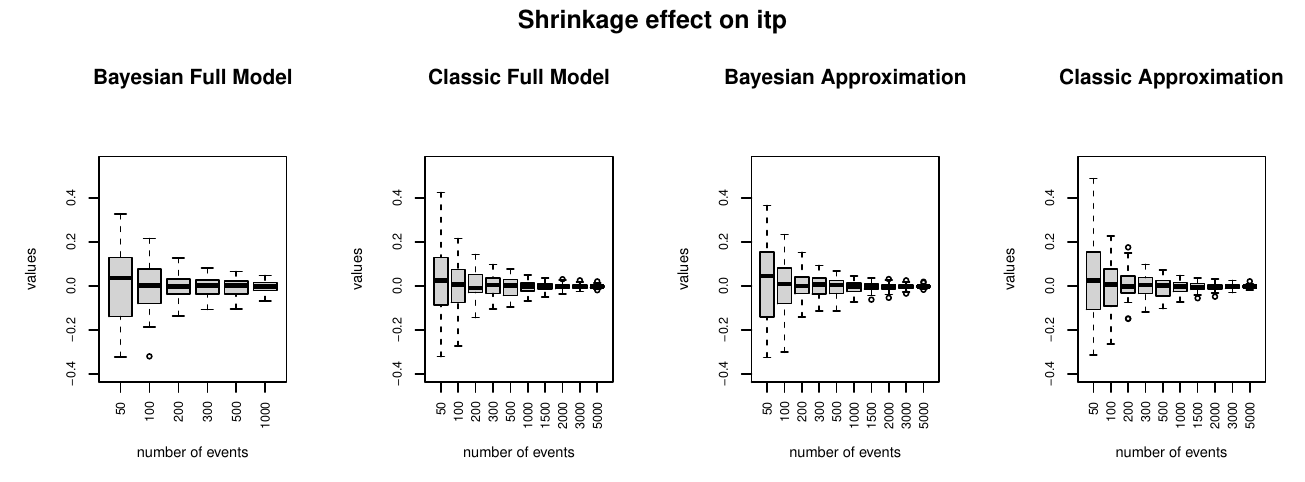}}\\ 
    \subfloat{\includegraphics[width = 6.5in]{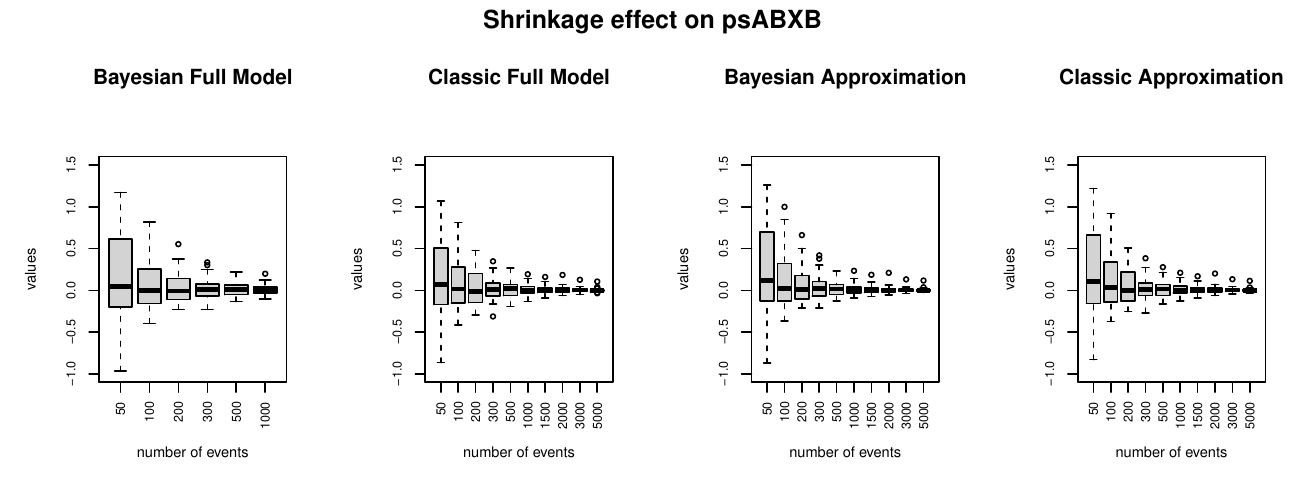}}\\ 
    }    
    \caption{Comparison of the shrinkage effect among the four multilevel models.}
    \label{fig:shrinkage}
\end{figure}

\begin{figure}
    \centering{
    \subfloat{\includegraphics[width = 6.5in]{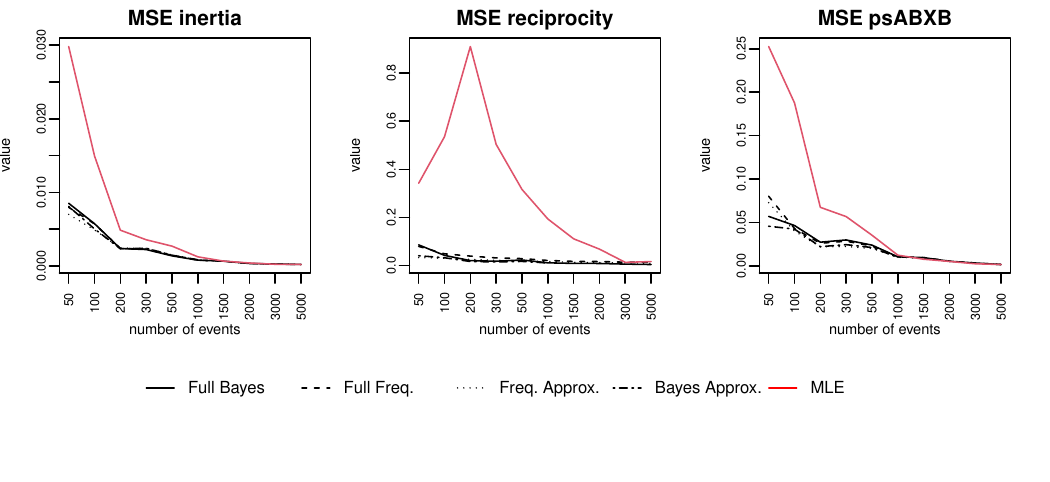}}
    }
    \caption{Comparison of mean squared error among multilevel estimator and independent estimator.}
    \label{fig:MSE}
\end{figure}

\paragraph{Estimation of random effects:} Shrinkage is defined as the difference between cluster-specific effects in the independent estimates (e.g. maximum likelihood estimates) and the estimates from the multilevel model. Thus, as the number of events (or the amount of information) in each cluster increases, we expect to see the shrinkage gradually decreasing towards zero. Figure \ref{fig:shrinkage} displays the results for the four models in three different covariates. As expected, the approximations produce a very similar degree of shrinkage as the exact models. For the efficiency of the estimator, we use the mean squared error (MSE) as a measure of the efficiency of the cluster-specific effect estimator. If $\hat{\bm{\beta}}$ is an estimator for $\bm{\beta}$, then the MSE is defined as $\text{MSE}(\hat{\bm{\beta}}) = E [{(\hat{\bm{\beta}} - \bm{\beta})}^2] = \text{Var}(\hat{\bm{\beta}}) - \text{Bias}^2(\hat{\bm{\beta}})$. Thus, if an estimator is unbiased, the MSE reduces its variance. Therefore, when comparing two estimators, the one with a smaller MSE is more efficient. Figure \ref{fig:MSE} shows the comparison between the MSE of the multilevel estimator for the four models and the independent estimator (MLE). It is clear that the efficiency of the estimator in the approximation is comparable to the exact model. Besides, we get some evidence of the advantage of the multilevel approach. The MSE of the multilevel estimator dominates the independent estimator (MLE) in all cases for small- to moderate-sized samples. In some cases (e.g. reciprocity), even when we have as many as 2000 events per network, there is still a considerable difference between the efficiency of the multilevel and the independent estimator.

\begin{figure}
    \centering{
    \subfloat{\includegraphics[width = 3.5in]{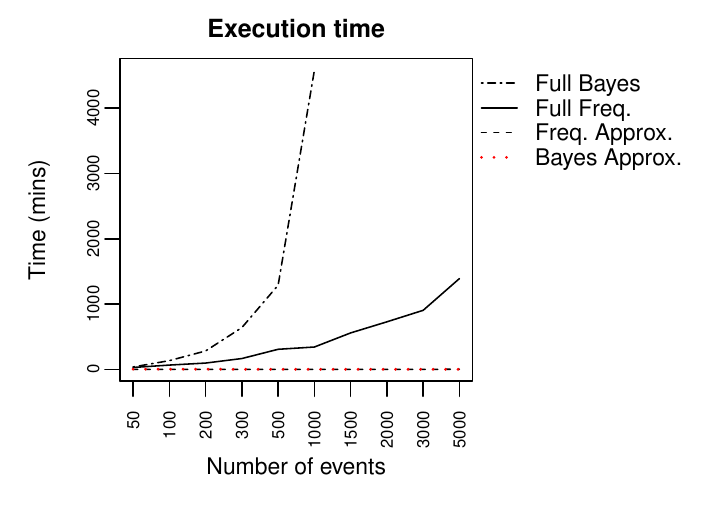}}\\ 
    }
    \caption{Comparison of running time for the four multilevel models.}
    \label{fig:run_time}
\end{figure}

\paragraph{Running time:} Figure \ref{fig:run_time} shows the comparison between running times for the four models across different sample sizes. The exact model takes longer in every single case. The exact Bayesian model, for example, displays a seemingly exponential growth in the running time. For 1000 events: the exact Bayesian model took 3 days and 4 hours to run; the exact frequentist model took 5 hours and 41 minutes; the Bayesian approximation took 31 seconds; and the frequentist approximation took 16 seconds. Thus, for large multilevel data, it does not seem feasible to run the exact model. Especially in the case when multiple models need to be compared, before deciding on a final model.


\section{Empirical applications}
\label{sec:applic}

\par
In this Section, all network statistics were computed using \textit{\textbf{remstats}} \citep{Meijerink2021}. The Bayesian approximations and the Bayesian exact models were estimated using \textit{\textbf{stan}} \citep{carpenter2017stan}. The frequentist approximations were fitted using \textit{\textbf{metafor}} \citep{viechtbauer2010} and \textit{\textbf{mixmeta}} \citep{gasparrini2019}. The frequentist exact models were estimated with 
\textit{\textbf{remstimate}} \citep{Arena2022} and \textit{\textbf{lme4}} \citep{bates2014fitting}. Besides, the experiments were run on a laptop with an Intel(R) Core(TM) i7-8665U processor with 4 cores and 16 GB RAM. 

\subsection{Data streams}

\par
In this experiment, we use the data set of email communications among employees of the defunct company Enron \citep{klimt2004introducing}. Following \cite{perry2013point}, only events with 5 receivers or less were included in the analysis. Those events with multiple receivers were broken up into dyadic interactions. The network that is analyzed consists of 153 actors and contains 32,261 events. The actors are divided across the `legal' department (25), the `trading' department (60), and `other' departments (68). They are almost evenly spread between Junior (79) and Senior (74) levels, and the majority are male (112).

\par
We emulate the streaming effect by starting with a small portion of the sequence and then gradually increase the number of events in batches. We start with $2000$ events and augment the sequence in batches of $1000$. The objective of this experiment is to illustrate the computational challenges of the exact model and to develop insights into to what extent the approximation models solve these issues. The model contains 20 covariates (inertia, reciprocity, psABAY, psABBY, psABXA, psABXB, out-degree sender, out-degree receiver, in-degree sender, in-degree receiver, outgoing two path, incoming two path, outgoing shared partner, incoming shared partners, recency rank receiver, same gender, same seniority, same department, same gender and difference department, different seniority and different department) plus an intercept. Thus, the full data matrix is a 3-dimensional object with dimensions $32,261 \times 23,256 \times 21$. We will evaluate the memory usage of both models (exact and approximation) and compare parameter estimates.

\begin{figure}
    \centering{
    \subfloat{\includegraphics[width = 3.5in]{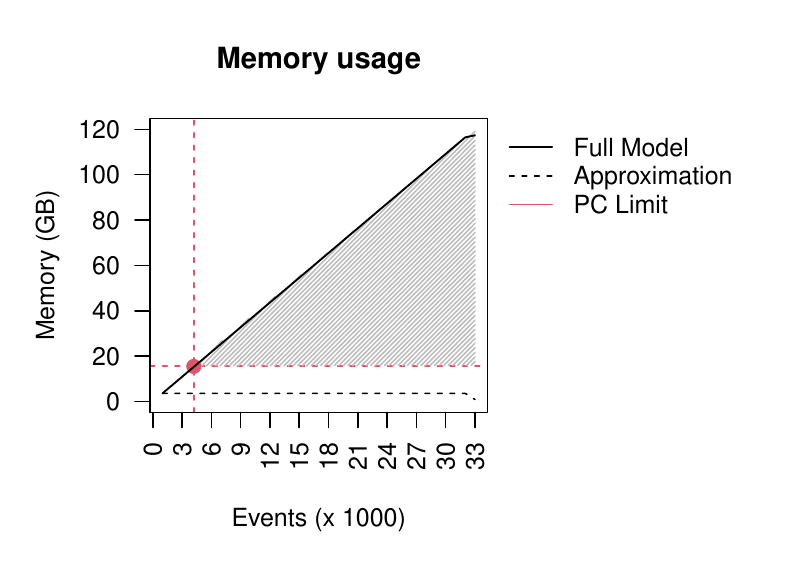}}\\ 
    }
    \caption{Memory usage of the exact model compared with the memory usage of the approximation.}
    \label{fig:memory_usage}
\end{figure}

\paragraph{Memory usage:} Figure \ref{fig:memory_usage} shows the differences in memory usage between the exact model and the approximation. The exact model displays a linear increase in the amount of memory used, whereas the approximation consumes about the same amount of memory overall. The red dot in the plot shows the memory limit of the laptop we used. For the exact model, after the second batch ($3,000$ events or less than $10\%$ of the total sequence), the R console issued the message "\textit{Error: cannot allocate vector of size 14.6 Gb}". Hence, the shaded area in figure \ref{fig:memory_usage}, represents a region where no resources are available to fit the exact model in the machine that we used to conduct our research. The approximation, however, fits in memory without problem.

\begin{figure}
    \centering{
    \subfloat{\includegraphics[width = 6.5in]{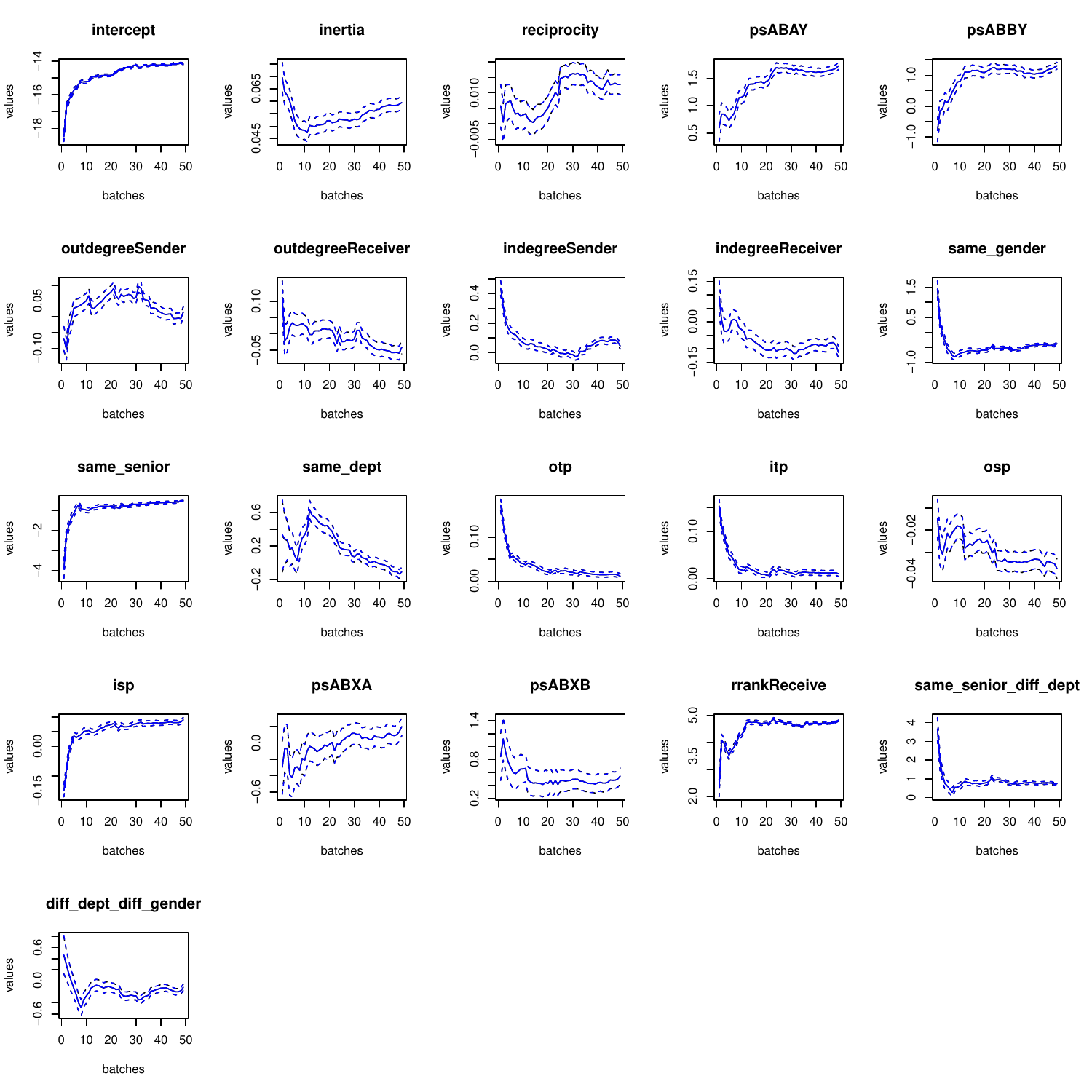}}\\ 
    }
    \caption{Comparison of the Bayesian (black lines) and frequentist (blue lines) approximations of the effects used in the application with the Enron data set. Solid lines are point estimates and traced lines are $95\%$ intervals.}
    \label{fig:enron_effect}
\end{figure}

\paragraph{Parameter estimates:} Figure \ref{fig:enron_effect} shows the evolution of the estimation along the event stream. Both models, Bayesian (black line) and frequentist (blue line), present virtually the same estimates to the point where it becomes impossible to distinguish them in the plot. One interesting aspect of Figure \ref{fig:enron_effect} is that for some effects, the behavior seen in the simulation study in Section \ref{sec:sim} is observed. The estimated effect starts a bit wiggly, and then it gets approximately constant, indicating some stability for that effect in the network. However, for some effects, the constant behavior is never observed, which suggests that those effects might be changing over time. Thus, this model also can also capture temporal variations in the effects that would simply be lost when the constant relational-event model is used. Table \ref{tab:enron_tab} shows parameters estimates, $95\%$ intervals for both models, and the widths of the intervals. These estimates are final. They contain information from all 32,261 events. Most estimates are identical, as already shown in the plots. Differences are seen only from the third decimal point. Finally, statistical significance is very similar between the models.

\subsection{Multilevel data}

\par
\begin{figure}
    \centering{
    \subfloat{\includegraphics[width = 3.5in]{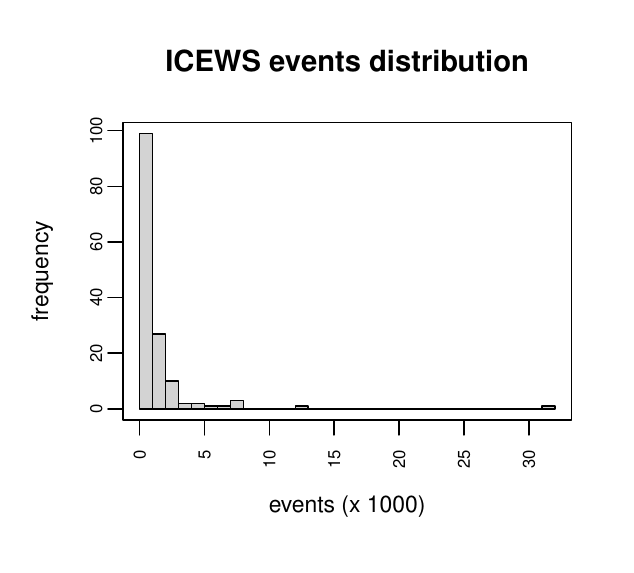}}\\ 
    }
    \caption{Distribution of events in the networks of the ICEWS data set.}
    \label{fig:icews_event_dist}
\end{figure}

To illustrate the multilevel models, we use the data set from Harvard dataverse Integrated Crisis Early Warning System (ICEWS) \citep{boschee2015}. These data consist of interactions between political actors collected from news articles. Here we consider the events in the year 2021. This data set consists of 147 countries where India has the largest sample with 31,513 events and Samoa has the smallest sample with just 62 events. Figure \ref{fig:icews_event_dist} shows that the distribution of events across those networks is highly skewed. 

\par
We fit a random effects model with 15 covariates (inertia, reciprocity, psABAY, psABBY, psABXY, psABXA, psABXB, out-degree sender, out-degree receiver,
in-degree sender, in-degree receiver, outgoing shared partners, incoming shared partners, recency rank sender and recency rank receiver) plus an intercept to these data. At first, an experiment is designed where the size of the data set is gradually increased. We varied number of networks, actors and covariates. Finally, the shrinkage effect and the parameter estimates for the full data set are analyzed. 

\paragraph{Data set size:} Table \ref{tab:rndICEWS} shows the results for simultaneously varying $K$, $N$ and $P$ in our model. Each model was left running for 12 hours to check if it can be fit within a reasonable amount of time. The (red crosses) green marks mean that the model (did not) run in time. The results show that the exact Bayesian model does not even run in time for a relatively small data set with $K = 50$, $N = 15$ and $P = 6$. The exact frequentist model fails to run within 12 hours for a moderately large data set with $K = 100$, $N = 25$ and $P = 11$. These results clearly show that for large multilevel data, it is computationally infeasible to run the exact model, regardless of whether the model is Bayesian or frequentist. The approximations, however, run on a fraction of the time assigned to all models. For $K = 147$, $N = 30$ and $P = 16$, the Bayesian approximation runs in approximately 1 hour, whereas the frequentist approximation runs in about 40 minutes. 

\begin{figure}
    \centering{
    \subfloat{\includegraphics[width = 4.5in]{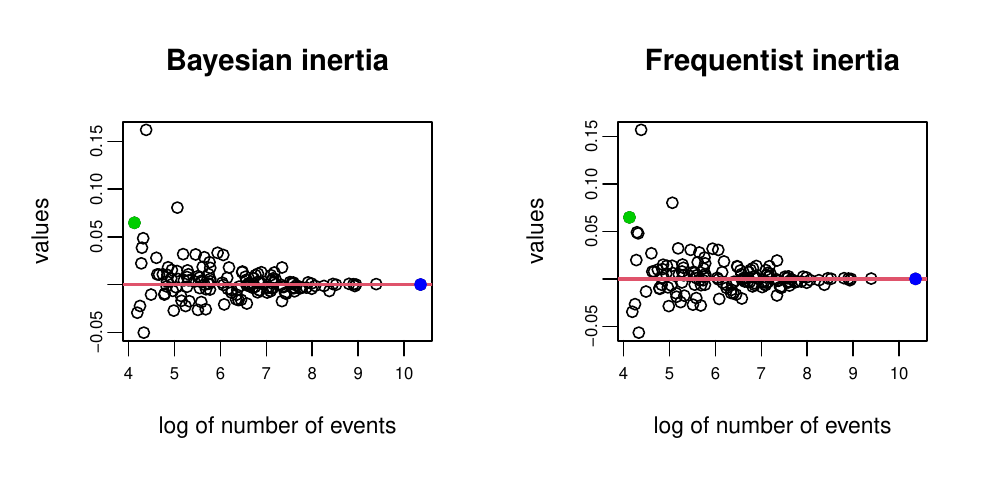}}\\ 
    \subfloat{\includegraphics[width = 4.5in]{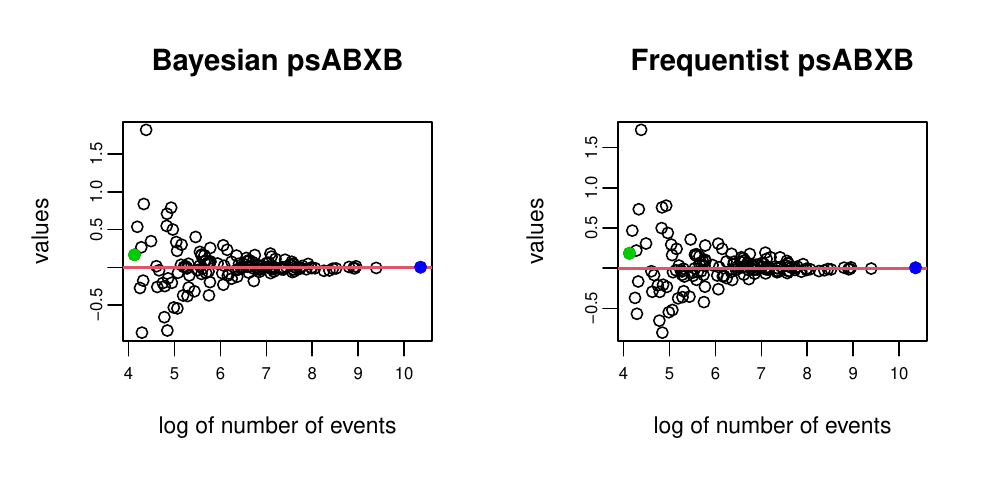}}\\ 
    \subfloat{\includegraphics[width = 4.5in]{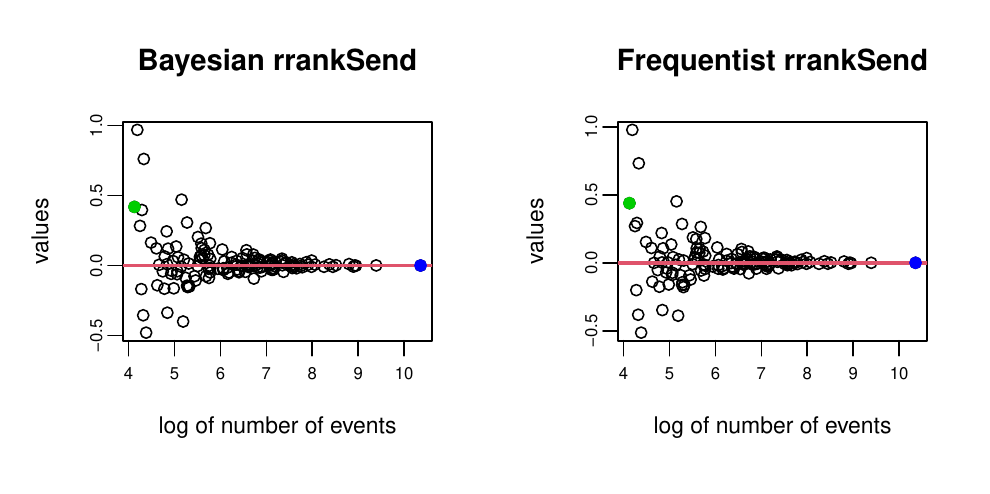}}\\ 
    }    
    \caption{Comparison of the shrinkage effect in the ICEWS data set for $K = 147$, $N = 30$ and $P = 16$. The green dot represents Samoa, the blue dot India and the red line represents no shrinkage.}
    \label{fig:ICEWSshrinkage}
\end{figure}

\paragraph{Shrinkage effect:} Shrinkage is the difference between the independent estimate (MLE) for the cluster-specific effect and the estimate provided by the multilevel model. The estimates presented were obtained using the complete data set and the largest model, $K = 147$, $N = 30$ and $P = 16$. Figure \ref{fig:ICEWSshrinkage} shows shrinkage results of the cluster-specific effect for a few covariates for both approximations. They exhibit very similar behaviors, with  smaller sample sizes showing larger shrinkage. The green dot represents Samoa, the blue dot India and the red line is zero, meaning that points lying on top of the line display no shrinkage. As expected the dot representing India, the largest network, always lies on top of the red line.

\paragraph{Parameter estimates:} Table \ref{tab:icews_rndEffects} displays estimates for the random effects means and variances for the Bayesian and frequentist approximations. The estimates were obtained for the largest setting $K = 147$, $N = 30$, and $P = 16$. As expected, the estimates from both models are very similar, both in terms of direction (positive or negative) and size. There is a slight difference in the intervals, which is due to the differences in model specification (since the Bayesian model has a Student's $t$ likelihood for the random effects). The participation shifts have all negative effects, which means that, on average they decrease the rates of communication. The largest positive effect is the recency rank sender, $\mu_{rrankSend} = 1.65$, which means that being the last one to send and event makes an actor more likely to send the next event, on average. Finally, most variance parameters are small, indicating low random effect variability. 


\section{Discussion and conclusion}
\label{sec:conclus}

\par
The analysis of relational event history data using relational event models has been plagued by computational issues due to memory storage limitations and computational complexity. This is, in particular, the case of a stream of relational events, multilevel (or clustered) relational event history data, but also in the case of large relational event history sequences.
In this paper, we introduced modeling approximations for relational event data streams and multilevel relational event data. The proposed approximations are based on statistical techniques that we borrow from the meta-analytic literature. In the case of data streams, newly observed batches of events are treated as new `studies'. In the case of multilevel relational event data, each independent relational event sequence is treated as a `study'. We then use meta-analysis to combine these `studies' to produce inferences. The relational event model for data streams is based on the assumption of constant effects which we aim to estimate as new batches of relational events are pouring in. Thus a fixed-effect meta-analytic model is employed for relational event data streams. Both classical and Bayesian methods were proposed for this purpose. The multilevel relational event model is based on the assumption of the existence of underlying heterogeneity among the independent networks, which makes a random-effect meta-analytic model the ideal approach. When all coefficients are assumed to be different across clusters, a classical random effects meta-analytic approximation was proposed. When certain coefficients are assumed to constant across clusters a Bayesian meta-analytic approximation was proposed using noninformative priors which can be used in a routine fashion. The goal of our approximations is to make the estimation of relation-event models feasible on standard desktop computers, which is currently not possible for the empirical relational event data that were considered in this paper. The algorithms developed have been implemented in the R-package \textit{\textbf{remx}} which is publicly available.

\par
For the data stream case, the network grows larger over time, and the interest is to update the inferences as new batches of events are observed. We provided a framework for these updates that rely only on newly observed events and that do not require doing any (re-)computations on the previously observed events. By avoiding the need to revisit past data points, the framework allows tremendously faster updating of the model and, at the same time, does not overload computer memory. We have shown that this approach approximates the relational-event model very well for network effects, but it seems to need large batches of events in order to properly approximate the model intercept. In addition, this model is also able to capture time variations in the covariate effects that would be lost when using the constant effect relational-event model.

\par
For the multilevel case, the number of networks and/or their sizes become too large to fit an exact multilevel model in a reasonable time, or the data no longer fit within the practical memory constraints of many computers. We alleviate these issues by using independent point estimates and uncertainty measures coming from each network separately as pseudo-data points. Then, we fit an approximated multilevel model that runs in a fraction of the time needed to fit a full multilevel relational event model. We showed that this simpler model behaves similarly to the exact model in terms of properties of the estimators, parameter recovery, and shrinkage behavior.

Although it has not been the focus of this paper, it is important to note that this fixed effects meta-analytic approximation can also be used for fitting relational event models to very large relational event sequences of, say, millions of events. Such large data sets are commonly observed in practice, but at the same time very problematic to analyze using the currently available (naive) approaches for relational event models. To use the proposed methodology, the big relational event sequence should be divided into batches, which are again treated as different `studies', similar to the case of data streams. Next a relational event model is fit to the separate batches, and the resulting fitted models are combined using the proposed meta-analytic approximation method. For future research, it would be interesting to see how this method compares to other previous to handle big relational event sequences using case-control sampling  \citep{lerner2013modeling,vu2015relational}. This comparison falls outside the scope of the current paper.

\par
Finally, other interesting future research directions include extending the proposed fast approximate methods to incorporate time-varying effects \citep{vu2011continuous, mulder2019modeling}. Moreover, the proposed meta-analysis approaches can be extended to actor-oriented relational event models \citep{stadtfeld2017dynamic, vieira2022}, which are helpful in separately modeling the behavioral choices on social interactions of senders and receivers.



\newpage

\appendix
\section{REM as Poisson regression model}
\label{app:poisson}

\par
For piece-wise constant exponential survival distributions, survival data can be modeled as Poisson regressions \citep{holford1980analysis, laird1981covariance}. This also applies to relational event data, facilitating the fitting of relational event models by means of available computer software for generalized (mixed-)linear models. Thus, we write the REM as a Poisson regression as follows

\begin{proof}
Assuming $M$ events are observed and the risk set contains $D$ dyads, 
\begin{align*}
    P (\textbf{E}\ |\ \bm{\beta}  ) &= \prod_{m = 1}^{M} P \Big((s_m, r_m, t_m) | \bm{\beta}, (e_{1}, \dots, e_{m-1}) \Big)  = \\
    &= \prod_{m = 1}^{M} \lambda_{s_m r_m}\ \times \exp \{-\Delta_{m} \sum_{(s,r) \in \mathcal{R}} \lambda_{s r}\} \\
    &= \prod_{m = 1}^{M} \exp \{\bm{\beta} \bm{x}'_{s_m r_m} \} \times \exp \{ - \exp \{\upsilon_{m} \} \sum_{(s,r) \in \mathcal{R}} \exp \{\bm{\beta} \bm{x}'_{s_m r_m} \}  \} &&\big(\upsilon_{m} = \log(\Delta_{m})\big)\\
    &\propto \prod_{m = 1}^{M} \exp \{\upsilon_{m} + \bm{\beta} \bm{x}'_{s_m r_m} \} \times \prod_{(s,r) \in \mathcal{R}} \exp \{ - \exp \{\upsilon_{m} + \bm{\beta} \bm{x}'_{s_m r_m} \}  \} \\
    &= \prod_{m = 1}^{M} \prod_{(s,r) \in \mathcal{R}} {\exp \{\upsilon_{m} + \bm{\beta} \bm{x}'_{s_m r_m} \}}^{y_{s r}} \times \exp \{ - \exp \{\upsilon_{m} + \bm{\beta} \bm{x}'_{s_m r_m} \}  \} \\
    &= \prod_{m = 1}^{M} \prod_{(s,r) \in \mathcal{R}} P \Big( y_{s_m r_m} | \lambda_{s_m r_m}= \exp \{\upsilon_{m} + \bm{\beta} \bm{x}'_{s_m r_m} \} \Big),
\end{align*}
where $y_{s_{m} r_{m}} = 1$ if dyad $(s, r)$ is observed and zero otherwise. As a result, the factorial term in the Poisson likelihood will always be equal to one, since $1! = 1$.\end{proof}

\par
Hence, by adding the inter-event time as offset in the event rate, the relational event model can be written as a Poisson regression model. An advantage of writing the relational event model as a Poisson regression is that we can handle relational event data for which the exact order of several events in shorter periods is unavailable. For example, relational event data between countries stored by digital news media are sometimes stored by only providing the events that occurred that day without reporting the exact timing or order within that day \citep{brandes2009networks}. Using a Poisson regression formulation, the observed dyads within each day are set to $y_{sr}=1$ while the other dyads are $y_{sr}=0$, so we do not need to use arbitrary event times that would cause bias. 

\par
In the multilevel setting, we multiply the likelihood of the independent event clusters by including a cluster-specific indicator,

\begin{equation}
\label{multilevel_poisson}
\begin{gathered}
    P (\textbf{E}\ |\ \bm{\beta}, \bm{\psi}  ) = \prod_{k = 1}^{K} \prod_{m = 1}^{M_{k}} \prod_{(s,r) \in \mathcal{R}_{k}} P ( y_{s_m r_m k} | \lambda_{s_m r_m k} = \exp \{\upsilon_{m k} + \bm{\psi} \bm{z}'_{s_m r_m k} + \bm{\beta}_{k} \bm{x}'_{s_m r_m k} \} )\\
   \bm{\beta}_{k} \sim \mathcal{N} (\bm{\mu}, \bm{\Sigma}),
\end{gathered}
\end{equation}

\noindent
where $M_{k}$ and $\mathcal{R}_{k}$ represent, respectively, the number of events and the risk set of cluster $k$.






\section{Estimates for the multilevel meta-analysis model}

The model is defined as

\begin{equation} 
\begin{gathered}
    \hat{\bm{\beta}}_{k}  \sim \mathcal{N} \Big(
    \bm{\mu}_{\beta} + \bm{\delta}_{k}, \hat{\bm{\Omega}}_{k} \Big)\\
    \bm{\delta}_{k} \sim \mathcal{N}(\bm{0}, \bm{\Sigma}).
\end{gathered}
\end{equation}

\noindent
The parameters to estimate are $\bm{\mu}_{\beta}$, $\bm{\delta}_{k}$ and $\bm{\Sigma}$. Then, assuming we have a model with $K$ networks and $p$ covariates, the likelihood function is given by

\begin{equation}
\begin{gathered}
    \mathcal{L}(\bm{\mu}_{\beta}, \bm{\delta}_{k}, \bm{\Sigma} | \hat{\bm{\beta}}_{k}, \hat{\bm{\Omega}}_{k}) = \prod_{k=1}^{K} \frac{1}{(2\pi)^{p/2}\ {|\hat{\bm{\Omega}}_{k}|}^{1/2} } \exp \Bigg(-\frac{1}{2} \Big( \hat{\bm{\beta}}_{k} - \bm{\mu}_{\beta} - \bm{\delta}_{k} \Big)'\ \hat{\bm{\Omega}}^{-1}_{k} \Big( \hat{\bm{\beta}}_{k} - \bm{\mu}_{\beta} - \bm{\delta}_{k}\Big) \Bigg) \times \\
    \times \frac{1}{(2\pi)^{p/2}\ {|\bm{\Sigma}|}^{1/2} } \exp \Bigg(-\frac{1}{2} \bm{\delta}'_{k}\ \bm{\Sigma}^{-1} \bm{\delta}_{k} \Bigg).
\end{gathered}
\end{equation}

\noindent
We need to take the logarithm and derive this function with respect to the parameters of interest in order to find the maximum likelihood estimates.

\paragraph{For $\bm{\delta}_{k}$:}

\begin{equation}
\begin{gathered}
    \frac{\partial}{\partial \bm{\delta}_{k}} = - \frac{1}{2} \Bigg( 2 \hat{\bm{\Omega}}^{-1}_{k} \Big( \hat{\bm{\beta}}_{k} - \bm{\mu}_{\beta}\Big) + 2 \bm{\delta}'_{k} \Big(\hat{\bm{\Omega}}^{-1}_{k} +  \bm{\Sigma}^{-1} \Big) \Bigg) = 0\\
    \bar{\bm{\delta}}_{k} = \Big(\hat{\bm{\Omega}}^{-1}_{k} +  \bm{\Sigma}^{-1} \Big)^{-1} \Big(\hat{\bm{\Omega}}^{-1}_{k} ( \hat{\bm{\beta}}_{k} - \bm{\mu}_{\beta}) \Big)
\end{gathered}
\end{equation}

\paragraph{For $\bm{\mu}_{\beta}$:}

\begin{equation}
\begin{gathered}
    \frac{\partial}{\partial \bm{\mu}_{\beta}} = - \frac{1}{2} \Bigg( 2 \sum_{k=1}^{K} \hat{\bm{\Omega}}^{-1}_{k} \Big( \hat{\bm{\beta}}_{k} - \bm{\delta}_{k}\Big) + 2 \bm{\mu}'_{\beta} \sum_{k=1}^{K} \hat{\bm{\Omega}}^{-1}_{k} \Bigg) = 0\\
    \bar{\bm{\mu}}_{\beta} = \Big(\sum_{k=1}^{K} \hat{\bm{\Omega}}^{-1}_{k} \Big)^{-1} \Big( \sum_{k=1}^{K} \hat{\bm{\Omega}}^{-1}_{k} ( \hat{\bm{\beta}}_{k} - \bm{\delta}_{k}) \Big)
\end{gathered}
\end{equation}

\paragraph{For $\bm{\Sigma}$:}

\begin{equation}
\begin{gathered}
    \frac{\partial}{\partial \bm{\Sigma}} = K \bm{\Sigma} - \sum_{k=1}^{K} \bm{\delta}_{k} \bm{\delta}'_{k} = 0 \\
    \bar{\bm{\Sigma}} = \frac{1}{K} \sum_{k=1}^{K} \bm{\delta}_{k} \bm{\delta}'_{k}
\end{gathered}
\end{equation}

\section{Meta-analytic approximation for multilevel relational event data}

Assuming with the maximum likelihood we are sampling observations from the distribution since $\bm{\theta}_{k}$ is the parameter vector in the sequence

\begin{equation}
    \bm{\hat{\theta}}_{k} \sim \mathcal{N}(\bm{\theta}_{k}, \bm{\hat{\Omega}}_{k})
\end{equation}

\noindent
Where $\hat{\bm{\theta}}_k$ is the estimated vector of parameters for sequence $k$, and $\bm{\theta}_{k}$ is the true vector of parameters for sequence $k$, which can contain fixed- and/or random-effects. Then, we'll use those observations as data to estimate the model as a model that takes into account dependencies between fixed-effects and random-effects means

\begin{equation}
    \hat{\bm{\theta}}_{k} = 
    \begin{bmatrix}
        \hat{\bm{\psi}}_{k} \\  
        \hat{\bm{\beta}}_{k} 
    \end{bmatrix} \sim \mathcal{N} \Bigg(
    \begin{bmatrix}
        \bm{\psi} \\
        \bm{\mu}_{\beta}
    \end{bmatrix} + 
    \begin{bmatrix}
        \bm{0} \\
        \bm{\delta}_{k}
    \end{bmatrix},
    \begin{bmatrix}
        \hat{\bm{\Omega}}_{k,11} \ \ \ \ \ \ \ \ \ \ \\
        \hat{\bm{\Omega}}_{k,21} \ \ \ \hat{\bm{\Omega}}_{k,22} 
    \end{bmatrix}
    \Bigg)
\end{equation}

\noindent
Where $\bm{\mu}_{\beta}$ refers to the mean of $\bm{\beta}_{k} = \bm{\mu}_{\beta} + \bm{\delta}_{k}$, and $\bm{\delta}_{k}$ are random-effects for group $k$.

\par
Next, to sample $\bm{\delta}_{k}$, we have that

\begin{equation}
    \begin{bmatrix}
        \bm{0} \\
        \bm{\delta}_{k}
    \end{bmatrix} \sim
    \mathcal{N} \Bigg(
    \begin{bmatrix}
        \hat{\bm{\psi}}_k - \bm{\psi}\\
        \hat{\bm{\beta}}_{k} - \bm{\mu}_{\beta}
    \end{bmatrix},
     \begin{bmatrix}
        \hat{\bm{\Omega}}_{k,11} \ \ \ \ \ \ \ \ \ \ \\
        \hat{\bm{\Omega}}_{k,21} \ \ \ \hat{\bm{\Omega}}_{k,22} 
    \end{bmatrix}
    \Bigg)
\end{equation}

\noindent
The prior for $\bm{\psi}$, $\bm{\mu}_{\beta}$ and the conditional distributions of $\bm{\delta}_{k}$ is given by

\begin{equation}
    \begin{gathered}
        \bm{\delta}_{k} \sim \mathcal{N} \Big((\hat{\bm{\beta}}_{k} - \bm{\mu}_{\beta}) + \hat{\bm{\Omega}}_{k,12} {\hat{\bm{\Omega}}}^{-1}_{k,11} (\bm{0} - (\hat{\bm{\psi}}_{k} - \bm{\psi})),  \hat{\bm{\Omega}}_{k,22} - \hat{\bm{\Omega}}_{k,21} {\hat{\bm{\Omega}}}^{-1}_{k,11} \hat{\bm{\Omega}}_{k,12}  \Big)\\
        \bm{\delta}_{k} \sim \mathcal{N} (\bm{0}, \bm{\Sigma}) \\
        p \Bigg(
        \begin{bmatrix}
           \bm{\psi} \\
           \bm{\mu}_{\beta} 
        \end{bmatrix}
        \Bigg) \sim 1\\
        \bm{\Sigma}\ |\ \textbf{A} \sim \mathcal{IW} \big(\eta + P - 1, 2 \eta \textbf{A} \big) \\
        \textbf{A} = \text{diag} \big(1/\alpha_{1}, 1/\alpha_{2}, \dots, 1/\alpha_{P} \big) \\
        \alpha_{i} \sim \mathcal{IG} \Big(\frac{1}{2}, \frac{1}{d^2_{i}} \Big),\ \text{for}\ i = 1, 2, \dots, P.
    \end{gathered}
\end{equation}

\noindent
For the prior of $\bm{\Sigma}$, check the manuscript. Let's make 

\begin{equation*}
    \begin{gathered}
        \hat{\bm{B}}_{k} = \hat{\bm{\Omega}}_{k,12} {\hat{\bm{\Omega}}}^{-1}_{k,11} (\bm{0} - (\hat{\bm{\psi}}_{k} - \bm{\psi}))\\
        \hat{\bm{S}}_{k} = \hat{\bm{\Omega}}_{k,22} - \hat{\bm{\Omega}}_{k,21} {\hat{\bm{\Omega}}}^{-1}_{k,11} \hat{\bm{\Omega}}_{k,12}.
    \end{gathered}
\end{equation*}

\noindent
The posterior distributions will be given by

\begin{equation}
    \begin{gathered}
       \begin{bmatrix}
          \bm{\psi} \\
          \bm{\mu}_{\beta}
       \end{bmatrix} |\ \hat{\bm{\beta}}, \bm{\beta}, \hat{\bm{\Omega}}, \bm{\delta}  \sim \mathcal{N} \Bigg( {\Big(\sum_{k} {\hat{\bm{\Omega}}}^{-1}_{k}\Big)}^{-1} \Big(\sum_{k} {\hat{\bm{\Omega}}}^{-1}_k (\hat{\bm{\theta}}_{k} - \bm{\delta}_{k}) \Big), {\Big(\sum_{k} {\hat{\bm{\Omega}}}^{-1}_{k}\Big)}^{-1} \Bigg)\\
       \bm{\delta}_{k} | \hat{\bm{\psi}}_{k}, \hat{\bm{\beta}}_{k}, \hat{\bm{\Omega}}_{k}, \bm{\psi}, \bm{\mu}_{\beta} \sim \mathcal{N} \Bigg( {\Big({\hat{\bm{S}}}^{-1}_{k} + {\bm{\Sigma}}^{-1}\Big)}^{-1} \Big( {\hat{\bm{S}}}^{-1}_{k}(\hat{\bm{\beta}}_{k} - \bm{\mu}_{\beta} + \hat{\bm{B}}_{k}) \Big), {\Big({\hat{\bm{S}}}^{-1}_{k} + {\bm{\Sigma}}^{-1}\Big)}^{-1} \Bigg)\\
       \bm\Sigma | \hat{\bm{\beta}}, \hat{\bm{\Omega}}, \bm{\delta}, \bm{\alpha} \sim \mathcal{IW} \Bigg( \eta + K + P - 1,\ 2 \eta \textbf{A} + \sum_{k=1}^{K} \bm{\delta}'_{k} \bm{\delta}_{k}  \Bigg)\\
       \alpha_{i} | \bm\Sigma \sim \mathcal{IG} \Bigg(\frac{\eta + P}{2}, \eta\ ({\bm{\Sigma}}^{-1})_{ii} + \frac{1}{d_{i}^2} \Bigg) 
    \end{gathered}
\end{equation}

\noindent
Where $\hat{\bm{\beta}}$, $\bm{\beta}$, $\bm{\delta}$, and $\hat{\bm{\Omega}}$ contain the parameters for all $k$, $({\bm{\Sigma}}^{-1})_{ii}$ is the $i^{th}$ diagonal value of ${\bm{\Sigma}}^{-1}$.


\newpage
\paragraph{Conflict of interest:}  On behalf of all authors, the corresponding author states that there is no conflict of interest.

\paragraph{Funding:} This research was supported by the Netherlands Organization for Scientific Research (NWO) Grant to JM and FV (452-17-006), as well as the ERC Starting Grant "TIMEISNOW" to JM and RL (758791).

\paragraph{Data Availability Statement:} The data associated with the manuscript is available upon request.

\bibliographystyle{apacite}
\bibliography{references}


\clearpage

\begin{center}
\begin{minipage}{.8\linewidth}
\begin{algorithm}[H]
\caption{Model for relational event data streams}\label{alg:meta_freq_single}
\KwData{$\{\textbf{E}_{1}, \textbf{E}_{2}, \dots, \textbf{E}_{L}\}$ and $\{\bm{X}_{1}, \bm{X}_{2}, \dots, \bm{X}_{L}\}$}
\KwResult{Vector of parameters $\bm{\beta}$}
  
  \For{$\ell \gets 1$ \KwTo $L$}{
  
    \If{$\ell > 1$}{
        Use $t_{_{M_{\ell}}}$ to correct the onset\;
    }
      $\textbf{E}_{\ell} \sim \text{REM}\big(\bm{\beta} | \bm{X}_{\ell}\big)$\;
      Store MLE $\hat{\bm{\beta}}_{\ell}$ and standard error $\hat{\bm{\Omega}}_{\ell}$\;

     \If{Frequentist approximation}{
            \eIf{$\ell = 1$} {
            
                $\tilde{\bm{\Omega}}(1) = \hat{\bm{\Omega}}_{1}$\;
                $\tilde{\bm{\beta}}(1) = \hat{\bm{\beta}}_{\ell}$\;
                
            } {
                $\tilde{\bm{\Omega}}({\ell}) = (\hat{\bm{\Omega}}({\ell}))^{-1} + \tilde{\bm{\Omega}}({\ell}-1))^{-1})^{-1}$\;
                $\tilde{\bm{\beta}}({\ell}) = \tilde{\bm{\Omega}}({\ell})\ (\hat{\bm{\Omega}}({\ell}))^{-1}\hat{\bm\beta}(\ell) +\tilde{\bm{\Omega}}({\ell}-1)^{-1} \tilde{\bm\beta}(\ell-1)))$\;
            }
    }
     \If{Bayesian approximation}{
            \eIf{$\ell = 1$}{
            
                $\bar{\bm{\Omega}}({1}) = \left(\hat{\bm{\Omega}}({1})^{-1}+\bm\Sigma_0^{-1}\right)^{-1}$\;
                $\bar{\bm{\beta}}({1}) = \bar{\bm{\Omega}}({1}) \left(\hat{\bm{\Omega}}({1})^{-1}\hat{\bm\beta}(1)+\bm\Sigma_0^{-1}\bm\beta_0\right)$\;
                
            } {
                $\bar{\bm{\Omega}}({\ell}) = \left(\hat{\bm{\Omega}}(\ell)^{-1}+ \bar{\bm{\Omega}}(\ell - 1)^{-1}\right)^{-1}$\;
                $\bar{\bm{\beta}}({\ell}) = \bar{\bm{\Omega}}(\ell)  \left(\hat{\bm{\Omega}}({\ell})^{-1}\hat{\bm\beta}(\ell)+\bar{\bm{\Omega}}(\ell - 1)^{-1} \bar{\bm\beta}(\ell - 1)\right)$\;
            }
    }
   } 

\end{algorithm}
\end{minipage}
\end{center}

\bigskip
\bigskip

\begin{center}
\begin{minipage}{.8\linewidth}
\begin{algorithm}[H]
\caption{Model for multilevel relational event data}\label{alg:multirem}
\KwData{\{$\textbf{E}_{k}$, $\bm{X}_{k}$\}, for $k = 1, \dots, K$.}
\KwResult{Parameters $\bm{\mu}$, $\bm{\Sigma}$, $\bm{\psi}$, and $\bm{\delta}_{k}$ $\forall\ k$}
  \For{$k\gets1$ \KwTo $K$}{
      $\textbf{E}_{k} \sim \text{REM}\big(\bm{\beta}_{k} | \bm{X}_{k}\big)$\;
      Store MLE $\hat{\bm{\theta}}_{k}$ and standard error $\hat{\bm{\Omega}}_{k}$\;
      
      \eIf{Model contain fixed effects}{
      
        Separate fixed effects\;
        $\hat{\bm{\psi}}_{k} = \hat{\bm{\theta}}_{k},\ \hat{\bm{\Psi}}_{k} = \hat{\bm{\Omega}}_{k,11} $\;
        $\hat{\bm{\beta}}_{k} = {\hat{\bm{\theta}}_{k},\ \hat{\bm{B}}_{k} = \hat{\bm{\Omega}}_{k,22}}$\;\
        Also keep $\hat{\bm{\Omega}}_{k, 21}$
      
      }{
       
        $\hat{\bm{\beta}}_{k} = \hat{\bm{\theta}}_{k}$\;
      
      }
      
    }
    
    \If{Frequentist approximation}{
    
    \For{$k \gets 1$ \KwTo $K$}{ 
    
      $\hat{\bm{\beta}}_{k} \sim \mathcal{N}(\bm{\mu}_{\beta} + \delta_{k}, \hat{\bm{\Omega}}_{k})$\;
      $\bm{\delta}_{k} \sim \mathcal{N}(\bm{0}, \bm{\Sigma})$
    }
    
    Optimize for $\bm{\delta}_{k}$, $\bm{\mu}_{\beta}$, and $\bm{\Sigma}$\;
    
    }
    
    \If{Bayesian approximation}{
    
        \While{MCMC does not reach convergence}{
            Set likelihood\;
        \For{$k\gets1$ \KwTo $K$}{
        $(\hat{\bm{\psi}}_{k}, \hat{\bm{\beta}}_{k}) \sim \mathcal{N}((\bm{\psi}, \bm{\mu}_{\beta} + \bm{\delta}_{k}), 
        \hat{\bm{\Omega}}_{k})$;\\
        $\bm{\delta}_{k}
        \sim \mathcal{N}(\bm{0}, \bm{\Sigma})$,;\
        }
        
        Set priors: $p(\bm{\delta}_{k})$, $p(\bm{\psi}, \bm{\mu})$, $p(\bm{\Sigma})$, $p(\bm{\alpha})$\;
    
        Obtain posterior sample from\;
        $[ \bm{\psi}, \bm{\mu}_{\beta} ]\ |\ \hat{\bm{\beta}}, \bm{\beta}, \hat{\bm{\Omega}}, \bm{\delta}$\;
        $\bm{\delta}_{k} | \hat{\bm{\psi}}_{k}, \hat{\bm{\beta}}_{k}, \hat{\bm{\Omega}}_{k}, \bm{\psi}, \bm{\mu}_{\beta}$\;
        $\bm\Sigma | \hat{\bm{\beta}}, \hat{\bm{\Omega}}, \bm{\delta}, \bm{\alpha}$\;
        $\alpha_{i} | \bm\Sigma\ \forall i$.
    }    
    
    }
    
\end{algorithm}
\end{minipage}
\end{center}

\clearpage

\clearpage

\clearpage

\clearpage

\clearpage

\clearpage

\clearpage

\clearpage

\clearpage

\begin{table}[ht]
\centering
\begin{tabular}{lccRc}
  \hline \hline
  \multicolumn{5}{c}{Approximations results for Enron data set}\\
  \hline
  & \multicolumn{2}{c}{Models} & \multicolumn{2}{c}{Interval width}\\
  \cline{2-3} \cline{4-5}
 Effect & Bayesian & Frequentist & Bayesian & Frequentist \\ 
  \cline{2-3} \cline{4-5}
  intercept & -14.163 (-14.213, -14.116) & -14.164 (-14.214, -14.114) & 0.098 & 0.1 \\ 
  inertia & 0.059 (0.057, 0.062) & 0.059 (0.057, 0.062) & 0.005 & 0.005 \\ 
  reciprocity & 0.013 (0.01, 0.016) & 0.013 (0.01, 0.016) & 0.006 & 0.006 \\ 
  psABAY & 1.726 (1.66, 1.791) & 1.725 (1.66, 1.791) & 0.131 & 0.132 \\ 
  psABBY & 1.305 (1.194, 1.419) & 1.307 (1.195, 1.42) & 0.225 & 0.225 \\ 
  outdegreeSender & 0.015 (-0.002, 0.031) & 0.015 (-0.002, 0.032) & 0.033 & 0.033 \\ 
  outdegreeReceiver & -0.041 (-0.061, -0.02) & -0.041 (-0.062, -0.02) & 0.041 & 0.042 \\ 
  indegreeSender & 0.046 (0.025, 0.066) & 0.045 (0.025, 0.066) & 0.041 & 0.041 \\ 
  indegreeReceiver & -0.114 (-0.136, -0.093) & -0.114 (-0.136, -0.092) & 0.043 & 0.043 \\ 
  same\_gender & -0.384 (-0.429, -0.34) & -0.385 (-0.432, -0.337) & 0.09 & 0.094 \\ 
  same\_senior & -0.5 (-0.561, -0.439) & -0.5 (-0.561, -0.439) & 0.122 & 0.122 \\ 
  same\_dept & -0.107 (-0.161, -0.054) & -0.106 (-0.158, -0.054) & 0.107 & 0.105 \\ 
  otp & 0.012 (0.006, 0.017) & 0.012 (0.007, 0.017) & 0.011 & 0.011 \\ 
  itp & 0.011 (0.005, 0.016) & 0.01 (0.005, 0.016) & 0.011 & 0.011 \\ 
  osp & -0.038 (-0.042, -0.033) & -0.038 (-0.042, -0.033) & 0.009 & 0.009 \\ 
  isp & 0.09 (0.082, 0.098) & 0.09 (0.082, 0.098) & 0.016 & 0.016 \\ 
  psABXA & 0.2 (0.09, 0.309) & 0.202 (0.09, 0.313) & 0.219 & 0.223 \\ 
  psABXB & 0.543 (0.418, 0.668) & 0.543 (0.418, 0.669) & 0.25 & 0.251 \\ 
  rrankReceive & 4.834 (4.794, 4.875) & 4.835 (4.795, 4.875) & 0.081 & 0.08 \\ 
  same\_senior\_diff\_dept & 0.737 (0.661, 0.816) & 0.738 (0.661, 0.815) & 0.154 & 0.154 \\ 
  diff\_dept\_diff\_gender & -0.117 (-0.173, -0.063) & -0.116 (-0.173, -0.06) & 0.11 & 0.114 \\ 
   \hline 
   \hline
\end{tabular}
\caption{\label{tab:enron_tab} Comparison of the Bayesian and frequentist approximations fitted to the Enron data set. These results are for the last batch, resulting on information from all the 32261 events. Numbers out of parentheses are point estimates (maximum likelihood estimate and posterior mean) and the numbers in the parentheses are $95\%$ intervals.}
\end{table}


\clearpage

\clearpage

\begin{table}[ht]
\centering
\begin{tabular}{lcccc}
\hline \hline
 \multicolumn{5}{c}{ICEWS data multilevel-model comparison by data set size}\\
  \hline
  & \multicolumn{2}{c}{Exact model} & \multicolumn{2}{c}{Approximation}\\
  \cline{2-5}
  Model size & Bayesian & Frequentist & Bayesian & Frequentist \\ 
  \hline
  K = 30, N = 10, P = 3 & \cmark & \cmark & \cmark & \cmark \\ 
  K = 50, N = 15, P = 6 & \xmark & \cmark & \cmark & \cmark \\ 
  K = 70, N = 20, P = 8 & \xmark & \cmark & \cmark & \cmark \\ 
  K = 100, N = 25, P = 11 & \xmark & \xmark & \cmark & \cmark \\ 
  K = 147, N = 30, P = 16 & \xmark & \xmark & \cmark &\cmark \\ 
   \hline
\end{tabular}
\caption{\label{tab:rndICEWS} Comparison of the four models according to model and data size. The models were fitted on the ICEWS data for the year 2021. The green check (red cross) means that the model run (does not run) before the time limit assigned runs out. As the data set gets larger, running the exact model becomes infeasible.}
\end{table}


\clearpage

\clearpage

\begin{table}[ht]
\centering
\begin{tabular}{lccRc}
  \hline \hline
  \multicolumn{5}{c}{Approximations comparison with estimates for ICEWS data set}\\
  \hline
  & \multicolumn{2}{c}{Bayesian} & \multicolumn{2}{c}{Frequentist}\\
  \cline{2-3} \cline{4-5}
 Effect & $\mu$ & $\sigma^2$ & \multicolumn{1}{c}{$\mu$} & $\sigma^2$ \\ 
  \cline{2-3} \cline{4-5}
intercept & -5.038 (-5.201, -5.145) & 0.972 (0.768, 1.229) & -5.033 (-5.191, -4.876) & 0.916 (0.711, 1.116) \\ 
  inertia & 0.023 (0.017, 0.019) & 0.001 (0.000, 0.002) & 0.023 (0.017, 0.029) & 0.001 (0.000, 0.002) \\ 
  reciprocity & -0.006 (-0.013, -0.01) & 0.001 (0.000, 0.002) & -0.007 (-0.013, -0.001) & 0.001 (0.000, 0.002) \\ 
  psABAY & -0.915 (-0.983, -0.959) & 0.106 (0.081, 0.141) & -0.923 (-0.991, -0.855) & 0.138 (0.107, 0.169) \\ 
  psABBY & -1.532 (-1.593, -1.573) & 0.082 (0.061, 0.109) & -1.54 (-1.604, -1.476) & 0.115 (0.089, 0.14) \\ 
  psABXY & -2.402 (-2.472, -2.448) & 0.144 (0.117, 0.178) & -2.408 (-2.484, -2.332) & 0.193 (0.15, 0.236) \\ 
  psABXA & -1.814 (-1.882, -1.857) & 0.095 (0.071, 0.123) & -1.824 (-1.891, -1.757) & 0.133 (0.103, 0.162) \\ 
  psABXB & -1.526 (-1.587, -1.567) & 0.077 (0.058, 0.101) & -1.532 (-1.592, -1.473) & 0.104 (0.08, 0.126) \\ 
  outdegreeSender & 0.284 (0.268, 0.273) & 0.007 (0.006, 0.010) & 0.283 (0.267, 0.300) & 0.008 (0.006, 0.009) \\ 
  outdegreeReceiver & 0.088 (0.07, 0.077) & 0.009 (0.007, 0.012) & 0.088 (0.071, 0.106) & 0.009 (0.007, 0.011) \\ 
  indegreeSender & 0.089 (0.073, 0.078) & 0.007 (0.005, 0.009) & 0.090 (0.074, 0.106) & 0.007 (0.005, 0.008) \\ 
  indegreeReceiver & 0.270 (0.255, 0.26) & 0.007 (0.005, 0.009) & 0.271 (0.256, 0.287) & 0.007 (0.005, 0.008) \\ 
  osp & 0.045 (0.036, 0.039) & 0.002 (0.002, 0.003) & 0.045 (0.036, 0.054) & 0.002 (0.002, 0.003) \\ 
  isp & 0.026 (0.017, 0.02) & 0.002 (0.001, 0.003) & 0.026 (0.018, 0.034) & 0.002 (0.001, 0.002) \\ 
  rrankReceive & 0.746 (0.688, 0.71) & 0.082 (0.06, 0.11) & 0.748 (0.696, 0.801) & 0.08 (0.062, 0.098) \\ 
  rrankSend & 1.655 (1.579, 1.605) & 0.195 (0.151, 0.252) & 1.65 (1.574, 1.725) & 0.189 (0.146, 0.23) \\ 
   \hline
\end{tabular}
\caption{\label{tab:icews_rndEffects} Estimates for random effects means and variances from Bayesian and frequentist approximations. These results are for the setting with $K = 147$, $N = 30$ and $P = 16$. Numbers out of parentheses are point estimates (maximum likelihood estimate and posterior mean) and the numbers in the parentheses are $95\%$ intervals.}
\end{table}


\end{document}